ORIGINAL RESEARCH

# Generalised probabilistic theories in a new light

## Raed Shaiia

The Center for Advanced Sciences (CAS), Al-Suwayda, Syria

**Correspondence**
Raed Shaiia, The Center for Advanced Sciences (CAS), Al-Suwayda, Syria.
Email: raedshaiia@gmail.com

**Abstract**
In this paper, a modified formulation of generalised probabilistic theories that will always give rise to the structure of Hilbert space of quantum mechanics, in any finite outcome space, is presented and the guidelines to how to extend this work to infinite dimensional Hilbert spaces are given. Moreover, this new formulation which will be called as extended operational-probabilistic theories, applies not only to quantum systems, but also equally well to classical systems, without violating Bell's theorem, and at the same time solves the measurement problem. A new answer to the question of why our universe is quantum mechanical rather than classical will be presented. Besides, this extended probability theory shows that it is non-determinacy, or to be more precise, the non-deterministic description of the universe, that makes the laws of physics the way they are. In addition, this paper shows that there is still a possibility that there might be a deterministic level from which our universe emerges, which if understood correctly, may open the door wide to applications in areas such as quantum computing. In addition, this paper explains the deep reason why complex Hilbert spaces in quantum mechanics are needed.

## 1 | INTRODUCTION

### 1.1 | Motivation

The motivation for this work was to deduce the Hilbert space structure of quantum mechanics from basic principles and to demystify some of the quantum phenomena to see whether that would help in the efforts to advance the research in quantum computing.

### 1.2 | Novelties

What this paper is not trying to do is to rephrase classical physics in the language of quantum mechanics, which is something that is well known to have been done in the literature. Rather, in this work we see that we can deduce the whole of quantum mechanics from a new formulation of ordinary probability theory alone (and maybe most profoundly, the deduction of the Schrödinger's equation itself). This work answers the question of why the laws of physics are the way they are without the need for a multiverse or for the anthropic principle; it establishes new relationships between space and time, explains why we need complex Hilbert spaces in quantum mechanics, and what connection this bears to the quantisation of space-time, gives new interpretation for the state vector and for some quantum phenomena such as the uncertainty principle and quantum entanglement, and it sheds a new light on the loopholes in Bell inequality tests of local realism, which in turn gives us new paths in the research to advance quantum computing.

### 1.3 | The Algorithm to this work

First, using category theory to rigorously build the needed tools and definitions.

Second, reformulating ordinary probability theory in the language of Hilbert spaces.

Third, deducing the whole of quantum mechanics from the previous structure, together with many new results.

This Algorithm is visualised using the next smart art figure:

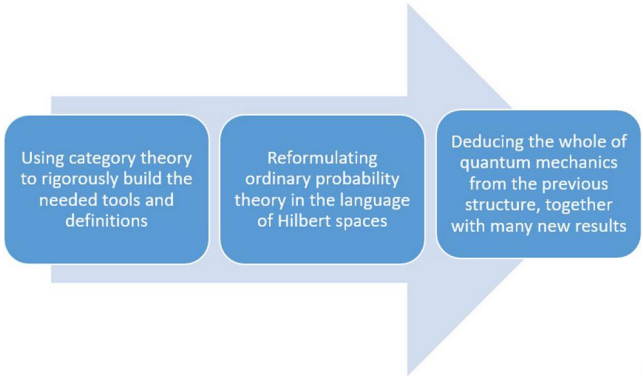

## 2 | DEFINITIONS

- A collection of sets and their subsets determine what we call a system. And the particular type of theory is what specifies which are the systems defined in it.

    For example, an elementary particle is described by a set of the possible values of its energy, a set of the possible values of its momentum, … etc.

    A coin is represented by a set of the possible outcomes of the experiment of throwing it, a singleton set that contains the value of its mass, … etc.

- Each collection of all the sets of some system that have the same cardinality defines what we call a state of the system.
- An event is an arrow connecting two sets of the system.

    An event of type $A \to B$ represents a possible transformation in the description of the system from a description in terms of $A$ to a description in terms of $B$.

$$\underline{\quad A \quad \boxed{\varepsilon} \quad B \quad} \qquad (1)$$

    The set to the left of the diagram is called *the input set of the event*, while the set to the right is called *the output set of the event*.

    For example, before throwing a coin, the sample space of the experiment done on the coin is $A = \{H, T\}$. But after we throw it and let us say it stabilises on $H$, the sample space of the experiment of reading the result will be $B = \{H\}$. In this case, the event of type $A \to B$ represents a possible transformation of the description of the coin which becomes a fact if we really throw the coin and it stabilises on $H$.

- An experiment on the system which measures one of its sets, let us say $A$, is a collection of events that have $A$ as their input set, and the union of their distinct output sets is $A$, such that each output set is a singleton of $A$.

    For example, the collection of two events of types $\{H, T\} \to \{H\}$ and $\{H, T\} \to \{T\}$ represents an experiment to measure $\{H, T\}$ if we were to throw the coin.

We notice that in this sense we are merely talking about a potential experiment up until now.

We can write the set $A$ as

$$A := \{a_i | i \in X\} \qquad (2)$$

where $X$ is an index set (we will call it from now on the outcome space) that indexes the elements of $A$, meaning $|X| = |A|$. Thus, we can represent the experiment as

$$\left\{ \underline{\quad A \quad \boxed{\varepsilon_i} \quad}^{\{a_i\}} \right\}_{i \in X} \qquad (3)$$

We say that each outcome space represents a property of the system, and each value of the index is a value of that property.

- When two events belong to the same experiment we say that they are coexisting.
- The sets that are the Cartesian product of the sets of two systems, together with their events composed in parallel (what we mean by that will be defined below), assign what we call a composite system of the two systems.

    For example, in the case of two coins, the set $\{H, T\} \times \{H, T\} = \{(H, H), (H, T), (T, H), (T, T)\}$ represents one of the sets that describe the composite system of the two coins.

- When the input and the output of an event represent a composite system, we draw boxes with multiple wires. For example, the box

$$\overset{A}{\underset{C}{\quad}} \boxed{\varepsilon} \overset{B}{\underset{D}{\quad}} \quad := \quad \overset{A \times C}{\quad} \boxed{\varepsilon} \overset{B \times D}{\quad} \qquad (4)$$

represents an event of type $(A \times C \to B \times D)$

### 2.1 | Composition of events

Events can be connected into networks through the following operations.

#### 2.1.1 | Sequential composition

An event of type $A \to B$ can be connected with an event of type $B \to C$, yielding an event of type $A \to C$.

#### 2.1.2 | Parallel composition

An event of type $A \to A'$ can be composed with an event of type $B \to B'$, yielding an event of type $(A \times B \to A' \times B')$

The sequential composition of two events $\varepsilon$ and $F$ of matching types is denoted by $F \circ \varepsilon$ and is represented graphically as

$$A \;\varepsilon\; B \;F\; C \quad := \quad A \;\boxed{F \circ \varepsilon}\; C \tag{5}$$

This graphical notation is justified by the requirement that sequential composition be associative, namely

$$G \circ (F \circ \varepsilon) = (G \circ F) \circ \varepsilon \tag{6}$$

for arbitrary events $\varepsilon$, $F$ and $G$. In addition to associativity, sequential composition is required to have an identity element for every set. The identity on set $A$, denoted by $I_A$, is the special event of type $A \to A$ identified by the conditions

$$A \;\boxed{I_A}\; A \;\boxed{\varepsilon}\; B \quad = \quad A \;\boxed{\varepsilon}\; B \tag{7}$$

and

$$B \;\boxed{F}\; A \;\boxed{I_A}\; A \quad = \quad B \;\boxed{F}\; A \tag{8}$$

required to be valid for arbitrary sets $A$, $B$ and arbitrary events $\varepsilon$ and $F$ of types $A \to B$ and $B \to A$, respectively. Consistently, we use the graphical notation

$$A \quad := \quad A \;\boxed{I_A}\; A \tag{9}$$

Mathematically, conditions (6), (7), and (8) impose the events form category [4], in which the sets are the objects and the events are the arrows.

Let us consider parallel composition. The parallel composition of two events $\varepsilon$ and $F$ is denoted as $\varepsilon \otimes F$ and is represented graphically as

$$A \;\boxed{\varepsilon}\; A' \;/\; B \;\boxed{F}\; B' \quad := \quad A\,B \;\boxed{\varepsilon \otimes F}\; A'\,B' \tag{10}$$

The graphical notation is justified by the requirement of the following condition

$$(\varepsilon \otimes F) \circ (G \otimes H) = (\varepsilon \circ G) \otimes (F \circ H) \tag{11}$$

where $\varepsilon$, $F$, $G$, and $H$ are arbitrary events.

In addition to Equations (7) and (8), parallel composition is required to satisfy the condition

$$I_{A \times B} = I_A \otimes I_B \tag{12}$$

We denote such category by Transf.

## 2.2 | Reversible events

An event $\varepsilon$ of type $A \to B$ is reversible if there exists another event $F$, of type $B \to A$, such that

$$A \;\boxed{\varepsilon}\; B \;\boxed{F}\; A \quad = \quad A \tag{13}$$

and

$$B \;\boxed{F}\; A \;\boxed{\varepsilon}\; B \quad = \quad B \tag{14}$$

When this is the case, we write $F = \varepsilon^{-1}$ and we say that sets $A$ and $B$ are operationally equivalent (or simply equivalent).

We denote by RevTransf $(A \to B)$ the set of reversible events of type $A \to B$. Such set (which may be empty) depends on the specific theory. In general, we require the existence of a reversible event that swaps pairs of sets. Given two sets $A$ and $B$, the swap of $A$ with $B$—denoted by $S_{A,B}$—is a reversible event of type $(A \times B) \to (B \times A)$, satisfying the condition

$$\boxed{S_{A,B}} \;\boxed{F}\; \;\boxed{\varepsilon}\; \boxed{S_{B',A'}} \quad = \quad \boxed{\varepsilon} \;\boxed{F} \tag{15}$$

for arbitrary sets $A$, $B$, $A'$, and $B'$ and arbitrary events $\varepsilon$ and $F$, as well as the conditions

$$\boxed{S_{A,B}} \;\boxed{S_{B,A}} \quad = \quad \begin{matrix} A \\ B \end{matrix} \tag{16}$$

and

$$\boxed{S_{A,B \times C}} \quad = \quad \boxed{S_{A,B}} \;\boxed{S_{A,C}} \tag{17}$$

## 2.3 | Summary about the operational structure

Summarising the ideas introduced so far, an operational structure consists of a triple

Op = (Transf, Outcomes, Experiments)

where Transf is a category, Outcomes is a collection of sets (the outcome spaces) closed under Cartesian product, and Experiments are sets of events as defined above.

## 3 | PROBABILISTIC STRUCTURE

In order to make predictions on the outcomes of the experiment, we need a rule assigning a probability to the events of such experiment. The rule is provided by the probabilistic structure of the theory, which we will give in a different manner than introduced in Ref. [1–3].

**Definition 1** (Probabilistic structure). Let Op be an operational structure. We say that Op is provided with a probabilistic structure if we can define on it for every experiment $\varepsilon$ that has an outcome space $X$, a map

$$P : \rho(X) \to [0, 1], \tag{18}$$

With $\rho(X)$ being the power set of $X$, and this map satisfies the following two requirements:

1.
$$P(X) = 1, \tag{19}$$

2. For any number of disjoint sets $A_1$, $A_2$, $A_3$,... that are subsets of $X$, we have:

$$P(U_i A_i) = \sum_i P(A_i) \tag{20}$$

whether this number is finite or infinite (in an infinite outcome space). The map $P$ need not be surjective: for example, in a deterministic theory the range of $P$ contains only the values 0 and 1.

From now on, we will use the following notation:

$$P(i) := P(\{i\}) \tag{21}$$

For any $i \in X$.

**Definition 2** An extended operational-probabilistic theory $\Theta$ is a pair (Op, P) consisting of an operational structure Op and of a probabilistic structure for Op as given above.

## 3.1 | Finite outcome spaces

### 3.1.1 | Experiments that have input sets with the same cardinality

In this section, we will take the cardinality of all outcome spaces of the experiments to be equal to the same positive integer $N$.

*Representing experiments by vectors*
From now on we will use Dirac notation for general vector spaces although these vector spaces need not be Hilbert spaces, unless we prove that.

Let us assume we have some experiment $\varepsilon$ with an outcome space $X$:

$$\varepsilon = \left\{ \begin{array}{c} A \\ \boxed{\varepsilon_i} \end{array} \!\! \begin{array}{c} \{a_i\} \end{array} \right\}_{i \in X} \tag{22}$$

where $|X| = N$.

Since $P$ is a probability function on $X$, then for any $B \subseteq X$ we have [5, 6]

$$P(B) = \sum_{i \in B} P(i), \qquad 0 \le P(i) \le 1 \tag{23}$$

Particularly,

$$P(X) = \sum_{i \in X} P(i) = 1 \tag{24}$$

On the other hand, in any inner-product vector space with a dimensionality $N$, where the inner product is positive-definite, and let us call this space $V$; we can always build an orthonormal basis for it using Gram–Schmidt theorem [7]. Let this basis be $\{|u_i\rangle\}_{i=1}^N$; then, for any vector $|C\rangle$ with a less than one square magnitude:

$$|C\rangle = \sum_{i=1}^N c_i |u_i\rangle \tag{25}$$

And

$$\langle C|C \rangle = \sum_{i=1}^N |c_i|^2, \qquad 0 \le |c_i|^2 \le 1 \tag{26}$$

In particular, for a normalised vector $|D\rangle$ we have

$$\langle D|D \rangle = \sum_{i=1}^N |c_i|^2 = 1 \tag{27}$$

Comparing Equations (23), (24), (26), and (27), we see that probability has the same behaviour as the square magnitude of a vector in $V$. This suggests another way for representing experiments, meaning, we can associate each output set of the experiment with a vector in such a vector space $V$, such that the squared-norm of the vector is equal to the probability of the output set. And to distinguish different experiments from each other, we will represent different experiments with different sets of vectors. As we said, we want each output set to be represented by a vector; this is why we need a set of $N$ different vectors to represent each experiment. We also want to represent different experiments by different sets of vectors. We can do that if we choose the dimensionality of the vector space to be equal to the cardinality of the input sets, meaning $N$, and represent each experiment by an ordered basis in $V$, such as $\{|u_i\rangle\}_{i=1}^N$ in a way that each output set is represented by a vector along one of the basis vectors with a squared-norm equal to its probability, with no two different experiments represented by the same basis.

Since we need the inner product to be positive-definite, that gives us two natural choices for the vector space $V$: either to choose it as a real vector space or as a complex vector space. For the moment, and for the sake of generality, we will choose it to be a complex vector space, then we will see later whether we can relieve this condition.

We will choose all the bases to be orthonormal, meaning, for any basis $\{|u_i\rangle\}_{i=1}^N$ we have

$$\langle u_i | u_j \rangle = \delta_{ij} \qquad (28)$$

Now we can define the function:

$$f : X \to V : i \mapsto c_i |u_i\rangle : |c_i|^2 = P(i), i \in X \qquad (29)$$

Actually, it is evident from the last definition that $f$ is not unique. In fact, there is an infinite number of functions satisfying the previous definition in a complex vector space [6].

Building on that, we can now define another function as follows:

$$g : \rho(X) \to V : Y \mapsto |Y\rangle = \sum_{i \in Y} f(i) = \sum_{i \in Y} c_i |u_i\rangle \qquad (30)$$

where $\rho(X)$ is the power set of $X$.

We immediately see that $g(\varnothing)$ is an empty sum, which means that the empty set is represented by the zero vector:

$$|\varnothing\rangle = 0 \qquad (31)$$

while $X$ itself will be represented by a normalised vector, and that is due to the fact that

$$|X\rangle = \sum_{i \in X} f(i) = \sum_{i \in X} c_i |u_i\rangle \qquad (32)$$

Thus,

$$\langle X | X \rangle = \sum_{i \in X} |c_i|^2 = \sum_{i \in X} P(i) = 1 \qquad (33)$$

where we can write the last equality due to the requirements of the probabilistic structure above.

In fact, as we see in Ref. [6], we can deduce from these definitions some very useful mathematical results. We will list some of them here without proofs, because their proofs can be found in Ref. [6].

We can see in Ref. [6] that for any two elements of $\rho(X)$, for example, $A$ and $B$, we have

$$P(A) = \langle A | A \rangle = \langle A | X \rangle = \langle X | A \rangle \qquad (34)$$

$$P(A \cap B) = \langle A \cap B | A \rangle = \langle A | A \cap B \rangle = \langle A \cap B | B \rangle \\ = \langle B | A \cap B \rangle = \langle A | B \rangle = \langle B | A \rangle \qquad (35)$$

$$A \cap B = \phi \Rightarrow |A\rangle \perp |B\rangle \qquad (36)$$

$$|A\rangle = |A \backslash B\rangle + |A \cap B\rangle \qquad (37)$$

$$|A \cup B\rangle = |A\rangle + |B\rangle - |A \cap B\rangle \qquad (38)$$

And if $|A\rangle = \sum_{r=1}^m a_r |u_r\rangle$ and $|B\rangle = \sum_{s=1}^l b_s |u_s\rangle$, then

$$|A \cap B\rangle = \sum_{r=1}^m \langle B | u_r \rangle | u_r \rangle = \sum_{s=1}^l \langle A | u_s \rangle | u_s \rangle \qquad (39)$$

And if we define $c(A) = X \backslash A$, then

$$|c(A)\rangle = |X\rangle - |A\rangle \qquad (40)$$

Since any finite dimensional inner-product vector space is a Hilbert space [8], we see immediately that this formulation of probability theory inevitably gives rise to Hilbert space structure.

## Observables

We will call any function from the events (labelled by outcomes) of an experiment to the real numbers an observable associated with that experiment, and we call the range of this function the *spectrum of this observable*. More specifically, if we have some experiment $\boldsymbol{\varepsilon} = \{\varepsilon_i\}_{i \in X}$, then an *observable associated with this experiment* is any function of the form

$$A : \boldsymbol{\varepsilon} = \{\varepsilon_i\}_{i \in X} \to \mathbb{R} : \varepsilon_i \mapsto a_i \qquad (41)$$

It is obvious that there is an infinite number of such functions [6].

And we have chosen the domain to be the experiment itself rather than the outcome space because what distinguishes an experiment is its events, not the outcome space.

We call any two observables that are associated with the same experiment compatible; otherwise, we call them incompatible.

If $A$ was injective, we say that its spectrum is non-degenerate. Otherwise, we say that it is degenerate.

We see that we can always build the following matrix out of the values of the spectrum of $A$:

$$\text{diag}(a_1, ..., a_N) \qquad (42)$$

which is a Hermitian matrix. Moreover, this matrix is the matrix representing the operator $\hat{A}$ in the vector space $V$, which satisfies

$$\hat{A}|u_i\rangle = a_i |u_i\rangle \qquad (43)$$

Since the matrix representing $\hat{A}$ is Hermitian, then $\hat{A}$ is a Hermitian operator, which as evident from the previous equation has the eigenvectors $\{|u_i\rangle\}_{i=1}^N$ that correspond to the eigenvalues $\{a_i\}_{i=1}^N$. So we can always associate any observable associated to an experiment with a Hermitian operator in which spectrum is the same as that of the observable, and its eigenvectors are the basis vectors that represent the experiment. This is why from now on when we say observable, we mean the Hermitian operator, unless otherwise stated. We notice that for this experiment we are talking about, all compatible observables share this set of basis vectors $\{|u_i\rangle\}_{i=1}^N$ as eigenvectors. That means they all commute, meaning, for any two compatible observables $\hat{A}$ and $\hat{B}$ we have [7]:

$$\left[\hat{A}, \hat{B}\right] = 0 \tag{44}$$

If we take any one of these observables, let it be $\hat{A}$, then we can think of the experiment as giving us one eigenvalue of this observable. And we will say that this observable is *measured using this experiment*.

Since this is true for every one of the observables compatible with $\hat{A}$ as we saw, then we will say that *compatible observables can be measured together with a single experiment*.

Since observables associated with different experiments will not have the same sets of eigenvectors, this means that the Hermitian operators representing them do not commute, and we say that *these observables cannot be measured simultaneously with the same experiment*.

Since we have chosen the experiment arbitrarily, the previous results hold for any experiment.

We see that the Heisenberg uncertainty principle is an inevitable result of this formulation. The only difference, which exists between the classical case and the quantum mechanical one, lies in the way we define the observables in each case and in which observables are compatible and which are not. And we will see later how Planck's constant enters the game when we return to the uncertainty principle with more detail later.

Now let us take a different experiment with an outcome space that has the same cardinality. We have said that we will represent it with a different basis. What is the relation between the bases representing different experiments?

Since all outcome spaces are represented by normalised vectors, which are linear combinations of the unit vectors representing their respective experiments, we will choose to represent all the outcome spaces by the same normalised vector, which has different components on different bases in $V$ representing different experiments, such that it gives right probabilities according to (29). The justification of the previous statement is the following lemma.

**Lemma** *In a finite dimensional inner-product complex vector space, we can always find an orthonormal basis such that we can give a specific unit vector, a set of components with desired squared magnitudes that are only subject to the condition that they sum into one.*

*Proof* First of all, from Gram-Schmidt theorem we know that in any inner-product vector space, we can build orthonormal bases [7].

Now, let $V$ be a finite inner-product space with a dimensionality $N$.

Suppose that the vector $|X\rangle$ has the set of components $\{c_1, \ldots, c_N\}$ on some orthonormal basis $\{|u_i\rangle\}_{i=1}^N$, And we want to find another orthonormal basis $\{|t_j\rangle\}_{j=1}^N$ in the space such that the vector $|X\rangle$ has the components $\{b_1, \ldots, b_N\}$ this new in basis, which satisfy.

$$|b_i|^2 = p_i, i \in \{1, \ldots, N\} \tag{45}$$

where

$$p_i \in [0, 1] \tag{46}$$

with

$$\sum_i p_i = 1 \tag{47}$$

let us assume the new basis vectors are given in terms of the old ones by

$$|t_j\rangle = \sum_{i=1}^N f_{ji} |u_i\rangle \tag{48}$$

Since the set $\{f_{ji}\}$ has $N^2$ elements for all the possible values of $i$ and $j$, and by noticing that each $f_{ji}$ is a complex number in general, thus it has two real numbers, this makes the number of unknowns that we need to find is $2N^2$.

Our task is then to find these unknowns. Let us count the number of equations we have.

Since the basis $\{|t_i\rangle\}_{i=1}^N$ is orthonormal, meaning:

$$\langle t_i | t_j \rangle = \delta_{ij} \tag{49}$$

This means we can write the components on it as

$$b_j = \langle t_i | X \rangle = \sum_j \langle t_i | u_j \rangle \langle u_j | X \rangle = \sum_j f_{ij}^* c_j \tag{50}$$

Which means that from equations (45) we see that we have $2N$ equations in the real unknowns. So to be able to find such a basis as the one we are looking for, the number of unknowns must be greater than or equals to the number of equations, meaning

$$2N^2 \geq 2N \Rightarrow N \geq 1 \tag{51}$$

Which is satisfied for any finite dimensionality of a vector space.

Notice that had we used a finite dimensional real vector space, the result would have been the same.

**Corollary** *Since we deal with probabilistic experiments with a sample space which has almost always two or more elements, the number of variables will almost always exceed the number of equations. Thus, we will have an infinite number of distinct orthonormal bases in which the vector representing sample space will have the right sets of components in all of them. This makes us able to represent even different experiments that have identical probability distributions, with different bases.*

From now on, we will call the normalised vector that represents all outcome spaces with the same cardinality *a state vector*, and we must stress that the same state vector represents all experiments with equal cardinality; this is why we will be saying *the state vector representing these experiments*.

And from now on, we say that all experiments that have outcome spaces with equal cardinality $N$ form a class $C_N$. And what we will call the state vector representing the class $C_N$ is the state vector representing experiments that their outcome spaces have the same cardinality $N$.

### Is the state vector unique?

We will replicate a derivation that is done in Ref. [6] just to make this paper self-contained. Is the state vector unique? In other words, can we use for a given class of experiments with the same cardinality more than one vector as a state vector to represent a certain state described by specific probability distributions for the possible experiments?

If it is not unique, then we must find the same probability distributions for all experiments of this class, whether we used $|X\rangle$ or—if exist—the other vector/vectors that can be used as state vectors.

Again, we will assume that the cardinality of outcome spaces of the experiments we are talking about is $N$.

Suppose that a state vector representing these experiments is $|X\rangle$.

We will take an experiment $\varepsilon_1$ and assume that the state vector is written using the basis representing the previous experiment according to (32) as

$$|X\rangle = \sum_{l=1}^{N} c_l |u_l\rangle \tag{52}$$

Now Let us take the vector $|X'\rangle$ which is

$$|X'\rangle = \sum_{l=1}^{N} c'_l |u_l\rangle : c'_l = c_l z_l \tag{53}$$

where $z_l$ are complex numbers which we will write in the form

$$z_l = A_l e^{i\theta_l} : A_l \in [0, \infty), \theta_l \in \mathbb{R} \tag{54}$$

And $i = \sqrt{-1}$

For $|X'\rangle$ to be a state vector, all the probabilities of each single outcome in the outcome space (thus all the probabilities of sets in the power set of the outcome space since the probability of such a set is equal to the sum of the probabilities of the outcomes that constitute it) for any experiment from this class must be the same as given by $|X\rangle$. So, the probabilities of the outcomes of $\varepsilon_1$ do not change.

So the following equation must hold

$$|c'_l|^2 = |c_l|^2 \tag{55}$$

Thus,

$$\begin{aligned} |A_l e^{i\theta_l} c_l|^2 &= |c_l|^2 \Rightarrow \\ |A_l|^2 |c_l|^2 &= |c_l|^2 \end{aligned} \tag{56}$$

And because the former condition is true even if we choose the experiment to satisfy $c_l \neq 0$ for all values of $c_l$, because our choice of $\varepsilon_1$ is arbitrary, we must have

$$|A_l|^2 = 1 \tag{57}$$

So we have the condition

$$A_l = 1 \tag{58}$$

which means that

$$z_l = e^{i\theta_l} \tag{59}$$

And that

$$c'_l = c_l z_l = c_l e^{i\theta_l} \tag{60}$$

But that is not enough, because the condition that probabilities must not change must be true for any other experiment of the same class and not just $\varepsilon_1$ because we are talking about state vectors that represent the same state here.

Let us take a different experiment $\varepsilon_2$ of the same class. We know that it must be represented by another basis, let us say $\left\{ |t_j\rangle \right\}_{j=1}^{N}$. We must have

$$|X\rangle = \sum_{j=1}^{N} b_j |t_j\rangle \tag{61}$$

We now have

$$|X\rangle = \sum_{j=1}^{N} b_j |t_j\rangle = \sum_{l=1}^{N} c_l |u_l\rangle \tag{62}$$

where

$$b_j = \langle t_j | X \rangle = \sum_{l=1}^{N} \langle t_j | u_l \rangle \langle u_l | X \rangle = \sum_{l=1}^{N} \langle t_j | u_l \rangle c_l \tag{63}$$

For $|X'\rangle$ we must have

$$|X'\rangle = \sum_{j=1}^{N} b'_j |t_j\rangle \tag{64}$$

So,

$$\begin{aligned} b'_j &= \langle t_j | X' \rangle = \sum_{l=1}^{N} \langle t_j | u_l \rangle \langle u_l | X' \rangle \\ &= \sum_{l=1}^{N} \langle t_j | u_l \rangle c'_l = \sum_{l=1}^{N} \langle t_j | u_l \rangle c_l e^{i\theta_l} \end{aligned} \tag{65}$$

We saw that the probabilities associated with the experiment $\varepsilon_1$ do not change. But to reach our goal, which is that we want $|X'\rangle$ to be a state vector too, then the probabilities associated with $\varepsilon_2$ must not change. So, we must have

$$|b'_j|^2 = |b_j|^2$$
$$b'_j b'^*_j = b_j b^*_j$$
$$\left( \sum_{l=1}^{N} \langle t_j | u_l \rangle c_l e^{i\theta_l} \right) \left( \sum_{k=1}^{N} \langle u_k | t_j \rangle c^*_k e^{-i\theta_k} \right)$$
$$= \left( \sum_{l=1}^{N} \langle t_j | u_l \rangle c_l \right) \left( \sum_{k=1}^{N} \langle u_k | t_j \rangle c^*_k \right)$$

So, we must have

$$\sum_{l,k} \langle t_j | u_l \rangle \langle u_k | t_j \rangle c_l c^*_k e^{i(\theta_l - \theta_k)} = \sum_{l,k} \langle t_j | u_l \rangle \langle u_k | t_j \rangle c_l c^*_k \tag{66}$$

The former equation must be true for any $c_k$ and $c_l$ because we are speaking of arbitrary experiments with arbitrary probability distributions. It is true when we fix the bases whatever $c_l$ were (we can fix the two bases and define an infinite number of experiments on them, meaning, whatever $c_l$ and $c^*_k$ were). So, their coefficients must be the same, which means

$$\begin{aligned} \langle t_j | u_l \rangle \langle u_k | t_j \rangle e^{i(\theta_l - \theta_k)} \\ = \langle t_j | u_l \rangle \langle u_k | t_j \rangle \Rightarrow \langle t_j | u_l \rangle \langle u_k | t_j \rangle \left[ e^{i(\theta_l - \theta_k)} - 1 \right] = 0 \end{aligned} \tag{67}$$

It must be true for all experiments, so for all bases even when $\langle t_j | u_l \rangle \neq 0$ for any $l$ and $j$. Thus,

$$e^{i(\theta_l - \theta_k)} = 1$$

which means that

$$e^{i\theta_l} = e^{i\theta_k} \Rightarrow \theta_l = \theta_k + 2\pi n, n \in \mathbb{Z} \tag{68}$$

And that is for any $l$ and $k$. So, we have

$$z_l = z_k \tag{69}$$

So, we see that

$$z_1 = z_2 = \ldots = z_N \tag{70}$$

And since all of them are pure phases, we can write

$$z_1 = z_2 = \ldots = z_N \equiv e^{i\theta}$$

So,

$$\begin{aligned} |X'\rangle &= \sum_{l=1}^{N} c_l e^{i\theta} |u_l\rangle = e^{i\theta} \sum_{l=1}^{N} c_l |u_l\rangle \Rightarrow \\ |X'\rangle &= e^{i\theta} |X\rangle \end{aligned} \tag{71}$$

So, for $|X'\rangle$ to be a state vector too, it must be of the former form. From the above we see that we can multiply $|X\rangle$ by any pure phase and still get another state vector. Immediately, we can see that if we were to work with a real vector space, the state vector would have been either $(|X\rangle)$ for $\theta = 0$ or $(-|X\rangle)$ for $\theta = \pi$.

### 3.1.2 | Coarse-grained measurements and mixed states

Since we have seen that the same mathematical structure of quantum mechanics is valid for any extended probability theory whether it describes classical or quantum systems, it will be helpful if we import some useful definitions from the mathematical structure of quantum mechanics to see what their interpretations are in the light of these new extended probabilistic theories.

Given vectors and dual vectors we can define operators of the form (in this section, we will follow to a large extent the steps taken in Ref. [9]):

$$|\varphi\rangle\langle\psi| \tag{72}$$

And if $|\psi\rangle$ is a state vector, the projection operator for this state is written as

$$\hat{P}_\psi = |\psi\rangle\langle\psi| \tag{73}$$

We will call it a density operator for a pure state.

We can even define a more general type of states, still described by density operators, by introducing 'mixtures' of pure states:

$$\hat{P} = \sum_{k=1}^{M} \lambda_k |\phi_k\rangle\langle\phi_k| \tag{74}$$

where $\{|\phi_k\rangle\}$ is some set of pure states, not necessarily orthogonal. The number $M$ could be anything and is not limited by the dimension of the Hilbert space. The $M$ numbers (or 'weights') $\lambda_k$ are non-zero and satisfy the relations:

$$0 < \lambda_k \le 1; \sum_{k=1}^{M} \lambda_k = 1 \tag{75}$$

The normalisation of the weights $p_k$ expresses the condition $\mathrm{Tr}\left(\hat{P}\right) = 1$.

Since $\hat{P}$ is Hermitian, we can diagonalise it, such that

$$\hat{P} = \sum_{k=1}^{N} p_k |\psi_k\rangle\langle\psi_k| \tag{76}$$

where the states $\{|\psi_k\rangle\}$ are orthogonal. The numbers $p_k$ satisfy

$$0 \le p_k \le 1; \sum_{k=1}^{N} p_k = 1 \tag{77}$$

The numbers $p_k$ are, in fact, nothing but the eigenvalues of $\hat{P}$. They sum to one because of normalisation. There are exactly $N = d$ of these numbers, where $d$ is the dimension of the Hilbert space.

Since this is the same familiar structure of quantum mechanics, we can use these two simple tests to determine whether $\hat{P}$ describes a pure state or not:

Pure state: $\hat{P}^2 = \hat{P}$; mixed state: $\hat{P}^2 \ne \hat{P}$.

Or,

Pure state: $\mathrm{Tr}\left[\hat{P}^2\right] = 1$; mixed state: $\mathrm{Tr}\left[\hat{P}^2\right] < 1$.

In fact, $P \equiv \mathrm{Tr}\left[\hat{P}^2\right]$ is called the purity of a state. A state is pure when its purity equals 1, and mixed otherwise.

We can certainly prepare a mixed state in a probabilistic way. If we prepare with probability $p_k$ a pure state $|\psi_k\rangle$, and then forget which pure state we prepared, the resulting mixed state is $\hat{P} = \sum_k p_k |\psi_k\rangle\langle\psi_k|$. In this case, $p_k$ certainly has the meaning of probability.

For example, if we have an experiment (for instance throwing an ordinary die) with a state vector,

$$|\psi\rangle = \sum_{k=1}^{6} \sqrt{p_k}|k\rangle, 0 \le p_k \le 1, \sum_{k=1}^{6} p_k = 1 \tag{78}$$

we can immediately see that it is a pure state:

$$\hat{P} = |\psi\rangle\langle\psi| \tag{79}$$

Because since $|\psi\rangle$ is normalised, we have

$$\hat{P}^2 = |\psi\rangle\langle\psi|\psi\rangle\langle\psi| = |\psi\rangle\langle\psi| = \hat{P} \tag{80}$$

But, if we take any other vector that represents a set which is a proper subset of the sample space, it will not be normalised; hence, we will get $\hat{P}^2 \ne \hat{P}$; thus, it represents a mixed state.

For example, in the previous die example, if we take the vector that represents having an odd number, it will be equal to

$$|\varphi\rangle = \sum_{n=1}^{3} \sqrt{p_{2n-1}}|2n-1\rangle \tag{81}$$

which will correspond to a mixed state. This vector gives us the probability to measure an odd number and this probability is $\langle\varphi|\varphi\rangle$ as we have seen in Ref. [6]. Have we done this measurement and know that we really got an odd number, then we would have done a coarse-grained measurement if this is all the information we got. So, mixed states can be used to represent coarse-graining measurements.

The density matrix operator in this case is (if the die is fair)

$$\hat{P} = \frac{1}{3}|1\rangle\langle1| + \frac{1}{3}|3\rangle\langle3| + \frac{1}{3}|5\rangle\langle5| \tag{82}$$

So, the density matrix is

$$\rho = \frac{1}{3}\mathrm{diag}(1, 0, 1, 0, 1, 0) \tag{83}$$

Thus,

$$\rho^2 = \frac{1}{9}\mathrm{diag}(1, 0, 1, 0, 1, 0) \tag{84}$$

Hence,

$$P = \mathrm{Tr}\left[\hat{P}^2\right] = \frac{1}{3} < 1 \tag{85}$$

### 3.1.3 | Entropy

Building on the previous die example, we see that if we have a vector that represents a subset of the sample space, and we write it in the form

$$|\psi\rangle = \sum_{n=1}^{M} c_n |u_n\rangle \qquad (86)$$

With $M \leq N$ where $N$ is the cardinality of the sample space, the density operator will be

$$\hat{P} = \sum_{n=1}^{M} p_n |u_n\rangle\langle u_n| = \sum_{n=1}^{M} |c_n|^2 |u_n\rangle\langle u_n| \qquad (87)$$

which means that the $ij$-element of the density matrix that represents this operator is

$$\rho_{ij} = p_i \delta_{ij} \qquad (88)$$

where the index $i$ runs over the basis vectors that appear in the expansion of $|\psi\rangle$ while the index $j$ runs over all the basis vectors.

Hence, the $ij$-element of the matrix that represents $\log\left[\hat{P}\right]$ is

$$\delta_{ij}\log p_i \qquad (89)$$

Thus, the $ij$-element of the matrix representing $\hat{P}\log\left[\hat{P}\right]$ will be

$$\delta_{ij} p_i \log p_i \qquad (90)$$

which means that

$$-\mathrm{Tr}\left(\hat{P}\log\left[\hat{P}\right]\right) = -\sum_{i=1}^{M} p_i \log p_i = S \qquad (91)$$

where $S$ is the entropy.

Now this means that the entropy of a pure state is zero, which corresponds to maximum knowledge, but it is a probabilistic knowledge about a class of probabilistic experiments that have the same number of outcomes as we saw. This means that if there is a deeper deterministic description of the situation, then there may be another knowledge hidden from us due to our probabilistic description that we can only get by knowing the deterministic description accurately.

### 3.1.4 | Composition of experiments

Not all collections of events are 'experiments'. Whether or not a specific collection is an experiment is determined by the theory, compatibly with the basic requirement that the set of experiments must be closed under sequential and parallel composition.

*Parallel composition*
The parallel composition of two experiments

$$\psi = \left\{ \boxed{A\; \boxed{\varepsilon_i}}^{\{a_i\}} \right\}_{i\in X} \quad , \quad \chi = \left\{ \boxed{B\; \boxed{F_j}}^{\{b_j\}} \right\}_{j\in Y} \qquad (92)$$

is defined to be

$$\psi \otimes \chi := \left\{ \boxed{\begin{array}{c} A\; \boxed{\varepsilon_i}^{\{a_i\}} \\ B\; \boxed{F_j}^{\{b_j\}} \end{array}} \right\}_{(i,j)\in X\times Y} = \left\{ \boxed{A\times B\; \boxed{\varepsilon_i \otimes F_j}}^{\{(a_i,b_j)\}} \right\}_{(i,j)\in X\times Y} \qquad (93)$$

and represents two possible non-deterministic processes occurring in parallel. The composition of experiments induces a composition of their outcome spaces via the Cartesian product. As a consequence, the set of all outcome spaces must be closed under this operation. We will denote such a set by Outcomes.

We see that the parallel composition can be interpreted as *an experiment* according to the definition we gave above for an experiment, and we will call this experiment *a composite experiment*.

For example, in the experiment of throwing two coins on parallel, if we assume that the experiment of throwing one coin is

$$F = \left\{ \boxed{\{H,T\}\; \boxed{\varepsilon_1}}^{\{H\}} \quad , \quad \boxed{\{H,T\}\; \boxed{\varepsilon_2}}^{\{T\}} \right\} \qquad (94)$$

then the composite experiment is

$$F \otimes F = \left\{ \boxed{\{H,T\}\; \boxed{\varepsilon_1}}^{\{H\}} \quad , \quad \boxed{\{H,T\}\; \boxed{\varepsilon_2}}^{\{T\}} \right\} \otimes \left\{ \boxed{\{H,T\}\; \boxed{\varepsilon_1}}^{\{H\}} \quad , \quad \boxed{\{H,T\}\; \boxed{\varepsilon_2}}^{\{T\}} \right\} \qquad (95)$$

$$= \left\{ \boxed{\{H,T\}^2\; \boxed{\varepsilon_1 \otimes \varepsilon_1}}^{\{(H,H)\}} \quad , \quad \boxed{\{H,T\}^2\; \boxed{\varepsilon_1 \otimes \varepsilon_2}}^{\{(H,T)\}} \quad , \quad \boxed{\{H,T\}^2\; \boxed{\varepsilon_2 \otimes \varepsilon_1}}^{\{(T,H)\}} \quad , \quad \boxed{\{H,T\}^2\; \boxed{\varepsilon_2 \otimes \varepsilon_2}}^{\{(T,T)\}} \right\}$$

## Sequential composition

First, we will explain the concept using the simple example of throwing a coin two consecutive times. Again, let us assume that the experiment of throwing a coin is

$$\boldsymbol{F} = \left\{ \underset{\varepsilon_1}{\overset{\{H,T\}}{\boxed{\varepsilon_1}}}^{\{H\}} \quad , \quad \underset{\varepsilon_2}{\overset{\{H,T\}}{\boxed{\varepsilon_2}}}^{\{T\}} \right\} \quad (96)$$

We can think of the composite experiment in this case as two experiments running in parallel: the experiment of throwing the coin in the first one while parallel to it, nothing happened in the second one and then the experiment of throwing the coin in the second one, while nothing changes regarding the first one:

In general, if we have the two experiments

$$\psi = \left\{ \overset{A}{\boxed{\varepsilon_i}}^{\{a_i\}} \right\}_{i \in X} \quad , \quad \chi = \left\{ \overset{B}{\boxed{F_j}}^{\{b_j\}} \right\}_{j \in Y} \quad (98)$$

then we define their sequential composition as

$$(97)$$

$$\chi \circ \psi := \left\{ \begin{array}{c} A \quad \xrightarrow{\{a_i\}} \varepsilon_i \xrightarrow{\{a_i\}} I_{\{a_i\}} \xrightarrow{\{a_i\}} \\ B \quad \xrightarrow{} I_B \xrightarrow{B} F_j \xrightarrow{\{b_j\}} \end{array} \right\}_{(i,j) \in X \times Y} = \left\{ \begin{array}{c} A \quad \xrightarrow{} \varepsilon_i \xrightarrow{\{a_i\}} \\ B \quad \xrightarrow{} F_j \xrightarrow{\{b_j\}} \end{array} \right\}_{(i,j) \in X \times Y}$$

$$= \left\{ \begin{array}{c} A \times B \\ \varepsilon_i \otimes F_j \end{array} \xrightarrow{\{(a_i, b_j)\}} \right\}_{(i,j) \in X \times Y} \tag{99}$$

$$= \psi \otimes \chi$$

Thus, we can see that also in the sequential composition we can get *an experiment* according to the definition we gave above for an experiment, and we will also call this experiment *a composite experiment*.

We can define a direction of time flow from top to bottom in the diagram, and we can define time steps using the number of identity events that appear in the previous equation, and in this case two, which in turn is the number of experiments we compose sequentially. Thus, we see that by default in this definition, the present cannot affect the past, and thus causality is inevitable.

We also see that we got the remarkable result:

$$\chi \circ \psi = \psi \otimes \chi \tag{100}$$

This is a remarkable result because it uncovers a strange and indeed deep correlation between space and time: it says that if for example, we throw a die in position $A$, and on parallel we throw a coin in position $B$, then this is equivalent from the point of view of category theory let us say to throwing the die, then throwing the coin afterwards. And if we exchange the initial positions of the coin and the die, then what is equivalent to that is throwing them in the reverse order in time. So, from category theory perspective, flipping the positions in space is equivalent to flipping the order in time.

Now we can see how this can remove some redundancy in probability theory: if we have two experiments whose sample spaces are $\Omega_1$ and $\Omega_2$, respectively, then we will have two options to represent the sample space of the composite experiment made out of the previous ones, and those options are either $\Omega = \Omega_1 \times \Omega_2$ or $\Omega' = \Omega_2 \times \Omega_1$. Thus, if we use one to represent parallel composition, then we can use the other to represent sequential composition.

### 3.1.5 | The representation of composite experiments in vector spaces

Let us take two experiments, where the first experiment $\varepsilon_1$ is from the class $C_N$ and the second experiment $\varepsilon_2$ is from the class $C_M$. Of course in general, these two experiments need not be on the same system.

We represent each experiment in its own vector space: $\varepsilon_1$ in $V_1$ and $\varepsilon_2$ in $V_2$.

Let us suppose we represent $\varepsilon_1$ in $V_1$ by

$$|X_1\rangle = \sum_{i=1}^{N} c_i |u_i\rangle \tag{101}$$

And that we represent $\varepsilon_2$ in $V_2$ by

$$|X_2\rangle = \sum_{j=1}^{M} b_j |t_j\rangle \tag{102}$$

We know from Section 3.1.4 that whether we composed the previous experiments sequentially or in parallel, the outcome space of the resulting experiment will be the Cartesian product of their outcome spaces. Thus, it will have the cardinality $N.M$; this is why we need to represent the composite experiment in a vector space with dimensionality $N.M$.

Since the vector space

$$V = V_1 \otimes V_2 \tag{103}$$

has this dimensionality, and so does the vector space $V' = V_2 \otimes V_1$, this means we can represent the composite experiment by a vector in one of them (we can, for example, represent the parallel composition in $V$ and the sequential composition in $V'$); let us call this vector $|X\rangle$. Let us say we represent the composite experiment in $V$. Since we can choose any orthonormal basis in $V$ to represent this experiment and since $\{|u_i t_j\rangle\}$ are orthonormal basis in this space, we can write

$$|X\rangle = \sum_{ij} f_{ij} |u_i t_j\rangle \in V_1 \otimes V_2 = V \tag{104}$$

where $\langle X|X\rangle = 1$ and $f_{ij} |u_i t_j\rangle$ representing $\{(u_i, t_j)\}$; thus,

$$p(u_i, t_j) = |f_{ij}|^2 \tag{105}$$

If

$$|X\rangle = |X_1\rangle \otimes |X_2\rangle, \tag{106}$$

then we call it a product state. Otherwise, we call it an entangled state.

A product state represents two non-interacting systems, while the opposite is true for the entangled state, as is extensively explained in Ref. [6].

We can easily generalise this to any finite number of experiments.

### 3.1.6 | Composite systems

From our definition in Section 2 of the composite system and from what we said about composing experiments, we see that we can treat an experiment on a composite system as a composite experiment on the subsystems that constitute the system and treat it according to the previous mathematical structure.

In particular, if the system is composed of two subsystems $A$ and $B$, then the states of the composite system are represented in the Hilbert space, which is the tensor product of the Hilbert spaces of the two subsystems $A$ and $B$:

$$H = H_A \otimes H_B$$

And to reflect the fact that the observables of the subsystem $A$ are defined to be the observables that act only in $H_A$ (and the same can be said regarding $B$), we can write them as

$$\hat{A}_i = \hat{\mathbf{A}}_i \otimes \mathbb{1}, \hat{B}_i = \mathbb{1} \otimes \hat{\mathbf{B}}_i$$

where the Hermition operators $\hat{\mathbf{A}}_i$ and $\hat{\mathbf{B}}_i$ act in $H_A$ and $H_B$, respectively.

We notice that we could have gotten for the experiment $\psi$, a number of results which is $N = |X|$ if we try to measure $X$.

Thus, the measurement for some property (which is the outcome space of some experiment) gives us one event of the experiment, corresponding to one value of the property.

Now, if we try to measure the *same property* again, meaning we want to measure $X$, thus to take the image of

$$\boldsymbol{M}_k(\boldsymbol{\psi}) = \left\{ \begin{array}{c} \{a_k\} \\ \boxed{I_{\{a_k\}}} \end{array}^{\{a_k\}} \right\} \tag{108}$$

then we can calculate the image only using the function $\boldsymbol{M}_k$ out of all the functions $\boldsymbol{M}_i$ where $i \in X$ because we have one event and one outcome in this experiment, and this outcome is $k$, and in this case, we will get

$$\boldsymbol{M}_k(\boldsymbol{M}_k(\psi)) = \boldsymbol{M}_k(\psi) \tag{109}$$

Thus, if we duplicate the same measurement, we will get the same result.

Now, let

$$\boldsymbol{\chi} = \left\{ \begin{array}{c} B \\ \boxed{F_j} \end{array}^{\{b_j\}} \right\}_{j \in Y} \tag{110}$$

be another experiment that we want to do after the measurement that gave us $k$. In this case, we will have

$$\boldsymbol{\chi} \circ \boldsymbol{M}_k(\boldsymbol{\psi}) = \boldsymbol{M}_k(\psi) \otimes \boldsymbol{\chi} = \left\{ \begin{array}{c} \{a_k\} \boxed{I_{\{a_k\}}}^{\{a_k\}} \\ B \boxed{F_j}^{\{b_j\}} \end{array} \right\}_{j \in Y} \quad = \quad \left\{ \begin{array}{c} \{a_k\} \times B \boxed{I_{\{a_k\}} \otimes F_j}^{\{(a_k, b_j)\}} \end{array} \right\}_{j \in Y} \tag{111}$$

### 3.1.7 | Measurement

We will denote the collection of all experiments by Experim. Then, we will define the following function and we can generalise this for any number of times.

$$\boldsymbol{M}_k : Experim \to Experim : \boldsymbol{\psi} = \left\{ \begin{array}{c} A \\ \boxed{\varepsilon_i} \end{array}^{\{a_i\}} \right\}_{i \in X} \longmapsto \left\{ \begin{array}{c} \{a_k\} \\ \boxed{I_{\{a_k\}}} \end{array}^{\{a_k\}} \right\} \tag{107}$$

for a specific value $k \in X$. We say that this function represents making a measurement of the property $X$ for the system under consideration and getting specifically as a result $k$.

### 3.1.8 | The measurement problem (observer-system composite system)

We will start by an example, then generalise.

Suppose we take the composite system of (coin-coin tosser). Let the experiment of tossing the coin be

$$\varepsilon = \left\{ \begin{array}{c} \{H,T\} \boxed{\varepsilon_1} \{H\} \end{array} \quad , \quad \begin{array}{c} \{H,T\} \boxed{\varepsilon_2} \{T\} \end{array} \right\} \tag{112}$$

While the experiment that represents what the tosser may see after tossing the coin is

$$O = \left\{ \begin{array}{c} \{O_H,O_T\} \boxed{F_1} \{O_H\} \end{array} \quad , \quad \begin{array}{c} \{O_H,O_T\} \boxed{F_2} \{O_T\} \end{array} \right\} \tag{113}$$

meaning that this observer may find the coin heads, which is represented by $O_H$, or the observer may find the coin tails, which is represented by $O_T$.

Now, after the toss, we cannot have simultaneously $H$ and $O_T$, or $T$ and $O_H$ because what the observer will see is defined by what the result of the toss is. Hence, the experiment that represents the potential change on the (coin/coin tosser) system will be

$$T = \left\{ \begin{array}{c} \{(H,O_H),(T,O_T)\} \boxed{T_1} \{(H,O_H)\} \end{array} \quad , \quad \begin{array}{c} \{(H,O_H),(T,O_T)\} \boxed{T_2} \{(T,O_T)\} \end{array} \right\} \tag{114}$$

We notice that we cannot get $T$ neither by parallel nor by sequential composition of $\varepsilon$ and $O$, which can be verified easily by simply taking their parallel and sequential compositions.

There is yet one more thing that we need to mention. Every time the coin is tossed and we get $H$, we simultaneously get $O_H$ for the observer, and $(H, O_H)$ for the composite system. The same is true if we replaced $H$ by $T$.

Thus, we must have

$$p(H, O_H) = p(O_H) = p(H) \tag{115}$$

and

$$p(T, O_T) = p(O_T) = p(T) \tag{116}$$

Now we will talk about the general case.
Suppose we have an experiment

$$\psi = \left\{ \begin{array}{c} A \boxed{\varepsilon_i} \{a_i\} \end{array} \right\}_{i \in X} \tag{117}$$

We say that the set O is one of the sets that describe an observer of the previous experiment if and only if the following conditions are met:

1. We can build the following two experiments that have the same outcome space as $\psi$:

$$O = \left\{ \begin{array}{c} O \boxed{F_i} \{O_i\} \end{array} \right\}_{i \in X} \tag{118}$$

and

$$\chi = \left\{ \begin{array}{c} \{(a_j,O_j)|j \in X\} \boxed{\chi_i} \{(a_i,O_i)\} \end{array} \right\}_{i \in X} \tag{119}$$

2. Using the probabilistic structures for the previous experiments, we have

$$p(a_i, O_i) = p(a_i) = p(O_i) \tag{120}$$

We say that $\chi$ represents the experiment done on the composite system of the observer and the original system that corresponds to $\psi$.

The state vectors of the three experiments, respectively, will be

$$|\psi\rangle = \sum_{i=1}^{N} c_i |a_i\rangle \in V_1 \tag{121}$$

$$|O\rangle = \sum_{i=1}^{N} b_i |O_i\rangle \in V_2 \tag{122}$$

$$|\chi\rangle = \sum_{i=1}^{N} d_i |a_i O_i\rangle \in V_1 \otimes V_2 \tag{123}$$

where the following condition must be satisfied:

$$|c_i|^2 = |b_i|^2 = |d_i|^2 \tag{124}$$

so that the probabilities are the same as we have mentioned above.

We immediately see that

$$|\chi\rangle \neq |\psi\rangle \otimes |O\rangle \tag{125}$$

So, the measurement is an entanglement between the system and the observer.

Furthermore, we notice that if we make a measurement and get the value $i$ of the property $X$, we will have the following three equations:

$$M_i(\psi) = \left\{ \frac{^{\{a_i\}}\boxed{I_{\{a_i\}}}^{\{a_i\}}}{} \right\} \tag{126}$$

$$M_i(O) = \left\{ \frac{^{\{O_i\}}\boxed{I_{\{O_i\}}}^{\{O_i\}}}{} \right\} \tag{127}$$

and

$$M_i(\chi) = \left\{ \frac{^{\{(a_i,O_i)\}}\boxed{I_{\{(a_i,O_i)\}}}^{\{(a_i,O_i)\}}}{} \right\}$$

$$= \left\{ \frac{^{\{a_i\}}\boxed{I_{\{a_i\}}}^{\{a_i\}}}{^{\{O_i\}}\boxed{I_{\{O_i\}}}^{\{O_i\}}} \right\}$$

$$= M_i(\psi) \otimes M_i(O) \tag{128}$$

$$= M_i(O) \circ M_i(\psi) \tag{129}$$

Now the question that arises is how to represent $M_k(\psi)$ by a vector. In fact, since the only outcome that is in the experiment $M_k$ is $k$, and we want this experiment to convey the meaning of measurement, and as we have seen according to (128), we can think of the experiments $M_i(\psi)$ and $M_i(O)$ as being done together, and once we have gotten an outcome for one of them, we get the corresponding outcome for the other; thus, they have the same probability distribution after doing the measurement, which is one for the outcome we got and zero for the others, and each experiment of them is a singleton; this is why we will represent them (up to a pure phase) by the following vectors, respectively:

$$|M_i(\psi)\rangle = \sum_{k=1}^{N} \delta_{ki}|a_k\rangle = |a_i\rangle \tag{130}$$

$$|M_i(O)\rangle = \sum_{k=1}^{N} \delta_{ki}|O_k\rangle = |O_i\rangle \tag{131}$$

And since we also get one outcome in the composite experiment after the measurement, we can here too write

$$|M_i(\chi)\rangle = \sum_{k=1}^{N} \delta_{ki}|a_k O_k\rangle = |a_i O_i\rangle \tag{132}$$

We can call this a collapse in the state vector. But we also see that there is nothing mysterious here, for we just have a change in probability distribution after the measurement.

Another way to express the above is, that if $\hat{A}$ is an observable that the experiment measures (as we have mentioned, that means a Hermitian operator that has $\{|a_i\rangle\}_{i=1}^{N}$ as its eigenvectors), then the system after the measurement will be in an eigenstate of $\hat{A}$ corresponding to the eigenvalue of it that we will measure [6].

Finally, the above work gives us a very important corollary:

We found that the state vector is merely a mathematical entity that represents the sample spaces of some class of experiments (with the meaning of a class of experiments we gave in section "Observables") that can be done on the system together with the probability distributions of the experiments of that class. This means that from a Bayesian probabilistic viewpoint, the state vector is subjective and can differ from observer to observer. But different descriptions of the system are due to different informational content that different observers have about the results of the experiments, since if they all had access to the information about the outcome of some experiment, all of their state vectors will collapse simultaneously. Hence, the collapse of the state vectors comes from doing the experiment itself and getting some definite outcome, with a knowledge about what outcome we got and not due to some observer observing the system per say. And it is for this reason that different observers get consistent results.

### 3.1.9 | Evolution of the states of a system

Let us assume that we have a system described by a state vector:

$$|\psi\rangle = \sum_{m} c_m |u_m\rangle \tag{133}$$

In some orthonormal basis $\{|u_m\rangle\}_{m=1}^{N}$. As we have seen in Section "Is the state vector unique?", any other vector of the form

$$|\varphi\rangle = \sum_{m} c_m e^{i\theta_m} |u_m\rangle \tag{134}$$

is another state vector for the system, which describes the same state only if all $\theta_m s$ are equal; otherwise, it will describe a state in which the probability distributions of at least some experiments are different.

This enables us to deal with two kinds of evolution for states: continuous evolution and discrete evolution.

#### Continuous evolution

If we start with a state that is given by Equation (134) and allow $\theta_m$ to vary continuously, then we can always write it as

$$\theta_m = \frac{g_m f_m(q^j)}{b_m} \tag{135}$$

Where $q^j$ are variables with $j = 1, \ldots, M$ for some positive integer $M$ and $b_m$ are some positive real constants that make $\theta_m$ dimensionless in case $g_m$ or $f_m$ or both have dimensions, and where $g_m$ are some real constants that depend only on $m$, and $f_m(q^j)$ are differentiable functions (hence, they are continuous). Of course by this, we are assuming that $\theta_m$ are differentiable functions.

We should mention that we chose to expand the state on the basis vectors of an experiment which the probabilities of its events do not change with $q^j$; of course, we can always find such experiments. For example, the experiments of reading some constants that characterise the system such as mass, electric charge, …etc. But since relative phases change with $q^j$, the probability distributions of other experiments will change in general.

As we have seen in section "Observables", we can always build a Hermitian operator $\hat{G}$ as follows:

$$\hat{G}|u_m\rangle = g_m|u_m\rangle \qquad (136)$$

Thus, we have

$$\frac{\partial|\varphi\rangle}{\partial q^j} = \frac{\partial}{\partial q^j}\sum_m c_m e^{-ig_m f_m/b_m}|u_m\rangle$$

$$= \sum_m -i\frac{g_m}{b_m}\frac{\partial f_m}{\partial q^j}c_m e^{-ig_m f_m/b_m}|u_m\rangle$$

$$= \sum_m -i\frac{1}{b_m}\frac{\partial f_m}{\partial q^j}c_m e^{-ig_m f_m/b_m}\hat{G}|u_m\rangle$$

$$= -i\hat{G}\sum_m \frac{1}{b_m}\frac{\partial f_m}{\partial q^j}c_m e^{-ig_m f_m/b_m}|u_m\rangle$$

And when we can choose a set of $q^j$ such that we can write

$$\theta_m = -\frac{g_m f(q^j)}{b} \qquad (137)$$

which is the case that we will focus on from now on, we will have:

$$\frac{\partial|\varphi\rangle}{\partial q^j} = -i\hat{G}\sum_m \frac{1}{b}\frac{\partial f}{\partial q^j}c_m e^{-ig_m f/b}|u_m\rangle$$

$$= -i\frac{\hat{G}}{b}\frac{\partial f}{\partial q^j}\sum_m c_m e^{-ig_m f/b}|u_m\rangle$$

$$= -i\frac{\hat{G}}{b}\frac{\partial f}{\partial q^j}|\varphi\rangle$$

Therefore,

$$\hat{G}_j|\varphi\rangle = ib\frac{\partial|\varphi\rangle}{\partial q^j} \qquad (138)$$

where we define the operators $\hat{G}_j$ as

$$\hat{G}_j = \frac{\partial f}{\partial q^j}\hat{G} \qquad (139)$$

We notice that since $q^j$ and $f(q^j)$ are real and $\hat{G}$ is Hermitian, $\hat{G}_j$ is also Hermitian. We can call $\hat{G}_j$ a generator for $q^j$ translation. In fact, Equation (138) is no other than Schrödinger's equation in its most general form.

Of course, we could have written $\theta_m$ using any other differentiable functions of other continuous variables, for example,

$$\theta_m = -\frac{l_m h(r^j)}{b}$$

with differentiable functions $h(r^j)$ and real constants $l_m$ that depend only on $m$, and a constant $b$ as before.

We can now build a new Hermitian operator $\hat{F}$ as follows:

$$\hat{F}|u_m\rangle = l_m|u_m\rangle \qquad (140)$$

and since $\hat{F}$ and $\hat{G}$ have the same eigenvectors, they commute.

Following the same procedure as before we find

$$\hat{F}_j|\varphi\rangle = ib\frac{\partial|\varphi\rangle}{\partial r^j} \qquad (141)$$

where we define the operators $\hat{F}_j = \frac{\partial h}{\partial r^j}\hat{F}$ and $\hat{F}_j$ (which will be Hermitian for similar reasons to the ones discussed in the case of $\hat{G}_j$) is a generator for $r^j$ translation.

Hence, we can start from any given state and follow its continuous evolution using these generators.

We must assert here that had we used real Hilbert spaces, then the only principal values that $\theta_m$ can take are 0 or $\pi$. Hence, we would have not been able to speak of any kind of continuous evolution of states at all. Thus for example, we would not have neither a continuous space nor a continuous time, because each one of them will enter as the continuous variable in the Schrödinger's equation above that has the right corresponding generator (the Hamiltonian in the case of time and the momentum in the case of space).

But we have to keep in mind that in this paper, we gave a very specific and new definition of the state vector: it is a vector that represents all the experiments with the same number of outcomes that can be done on the system, but not all the experiments of any number of outcomes that can be done on the system, as we have seen in section "Observables". And we have proved that we cannot have a continuous evolution for this vector unless we use complex Hilbert spaces. And since this paper shows that it is this vector that we usually call the state vector in ordinary quantum mechanics, since it is the vector used to get from it the probabilities of measurement outcomes according to the Born rule, as explained in sections "Representing experiments by vectors" and "Observables", together with the fact that its continuous evolution is governed by the Schrodinger's equation as we have shown in this section, thus comes the result that we cannot have continuous evolution of states unless we used complex Hilbert spaces.

Now, if we choose a 'time variable', meaning, a real variable that all other dynamical variables (such as $q^j$) depend on, and using Einstein's summation convention, where two repeated indices, one downstairs and the other upstairs, are summed over, we can write the following:

$$i \frac{\partial |\varphi\rangle}{\partial t} = i \frac{\partial |\varphi\rangle}{\partial q^j} \frac{\partial q^j}{\partial t}$$

Using Equation (138):

$$i \frac{\partial |\varphi\rangle}{\partial t} = \frac{\hat{G}_j |\varphi\rangle}{b} \frac{\partial q^j}{\partial t}$$

Hence,

$$\frac{\partial q^j}{\partial t} \hat{G}_j |\varphi\rangle = ib \frac{\partial |\varphi\rangle}{\partial t}$$

We will define the Hamiltonian to be

$$\hat{H} := \frac{\partial q^j}{\partial t} \hat{G}_j \qquad (142)$$

which means that we can write

$$\hat{H} |\varphi\rangle = ib \frac{\partial |\varphi\rangle}{\partial t}$$

Or since all the dynamical variables depend on time, we can write it as

$$\hat{H} |\varphi\rangle = ib \frac{d |\varphi\rangle}{dt} \qquad (143)$$

Of course, since $q^j$ and $t$ are real, and $\hat{G}_j$ is Hermitian for any $j$, then $\hat{H}$ is a Hermitian operator. And we will see later in the example below how to build Hamiltonians.

Using Equation (139) and Equation (142) we find that

$$\left[\hat{H}, \hat{G}_j\right] = \left[\hat{H}, \hat{G}\right] = \left[\hat{G}_j, \hat{G}\right] = \left[\hat{G}_j, \hat{G}_k\right] = \hat{0} \qquad (144)$$

which means that the experiment that is represented by the basis $\{|u_m\rangle\}$ measures all the observables $\hat{H}, \hat{G}_j$ and $\hat{G}$. We will call the eigenvalues of $\hat{H}$ the energies of the system.

From Equation (137), we see that the dimensions of $b$ that make $\theta_m$ dimensionless as follows:

$$[b] = [g_m] \left[f\left(q^j\right)\right]$$

And from Equation (139),

$$\left[G_j\right] = \frac{\left[f\left(q^j\right)\right] [g_m]}{[q^j]}$$

While from Equation (142) we have

$$[H] = \frac{[q^j] [G_j]}{[T]}$$

which means using the previous three equations that

$$[b] = [H][T]$$

This means that the dimensions of the constant $b$ are energy multiplied by time, which is amazing because those are the dimensions of Planck's constant.

Of course, $b$ itself is nothing but Planck's constant. And in a fundamental model like this, what is important is to deduce the dimensions of the constant relative to other physical quantities and not its numerical value because the said numerical value depends on the arbitrary choice of units, but the said dimensions will remain the same. For example, the area of a $1 \ m^2$ square can be written as $10^4 \ cm^2$, depending on our arbitrary choice of unites. But in both cases, the dimensions of the area are $[L]^2$. And since the constant $b$ enters in any probabilistic description, whether for atoms or for macroscopic objects, then using units defined based on macroscopic quantities (quantities that describe large numbers of atoms), we are guaranteed to get a very small value for that constant in those units, as is really the case.

Of course, throughout this paper I will keep using the symbol $b$ for this constant rather than $\hbar$ to stress the fact that this description does not make any differentiation whether we used it for macroscopic or microscopic systems (traditionally classical or quantum).

From Equation (142) and Equation (139) we find

$$\hat{H} = \frac{dq^j}{dt} \hat{G}_j = \frac{dq^j}{dt} \frac{\partial f}{\partial q^j} \hat{G} \Rightarrow \hat{H} = \frac{df}{dt} \hat{G}$$

While from Equation (136),

$$\hat{G} |u_m\rangle = g_m |u_m\rangle \Rightarrow$$

$$\frac{df}{dt} \hat{G} |u_m\rangle = \frac{df}{dt} g_m |u_m\rangle \Rightarrow$$

$$\hat{H} |u_m\rangle = \frac{df}{dt} g_m |u_m\rangle$$

And since we defined the energies to be the eigenvalues of the Hamiltonian, meaning

$$\hat{H} |\psi\rangle := E_m |\psi\rangle \Rightarrow E_m = \frac{df}{dt} g_m$$

This means that since

$$\frac{df}{dt} = \frac{E_m}{g_m}$$

then it must be the case that

$$\frac{E_m}{g_m} = constant$$

where what we mean by constant is a constant with respect to $m$.

Now,

$$\hat{G}|u_m\rangle = g_m|u_m\rangle \Rightarrow$$

$$\frac{\partial f}{\partial q^j}\hat{G}|u_m\rangle = g_m\frac{\partial f}{\partial q^j}|u_m\rangle \Rightarrow$$

$$\hat{G}_j|u_m\rangle = g_m\frac{\partial f}{\partial q^j}|u_m\rangle$$

This means that the eigenvalues of $\hat{G}_j$ are given by $g_m\frac{\partial f}{\partial q^j}$. From section "Observables", we see that neither $E_m$ nor $g_m$ depend on time because they are the images of the events of the experiment represented by the basis $\{|u_m\rangle\}$ under the functions that define the observables: $H$ and $G$. And each event can only have one image under a function. This means, since we found

$$\frac{df}{dt} = \frac{E_m}{g_m}$$

that

$$f = \frac{E_m}{g_m}t + A$$

where $A$ is a constant that do not depend on time.

But according to Equation (134) and Equation (137) we can write

$$|\psi\rangle = \sum_m c_m e^{-\frac{ig_m f}{\hbar}}|u_m\rangle$$

So, substituting the expression of $f$ in the previous equation we find

$$|\psi(t)\rangle = \sum_m c_m e^{-\frac{ig_m}{\hbar}\left(\frac{E_m t}{g_m} + A\right)}|u_m\rangle$$

Or

$$|\psi(t)\rangle = \sum_m c_m e^{-\frac{ig_m A}{\hbar}}e^{-i\frac{E_m t}{\hbar}}|u_m\rangle$$

from which we immediately find

$$|\psi(0)\rangle = \sum_m c_m e^{-\frac{ig_m A}{\hbar}}|u_m\rangle$$

If we define the wave function as

$$\psi_n(t) = \langle u_n|\psi(t)\rangle$$

then

$$\langle u_n|\psi(t)\rangle = \sum_m c_m e^{-\frac{ig_m A}{\hbar}}e^{-i\frac{E_m t}{\hbar}}\delta_{nm} \Rightarrow$$

$$\psi_n(t) = c_n e^{-\frac{ig_n A}{\hbar}}e^{-i\frac{E_n t}{\hbar}} \Rightarrow$$

$$\psi_n(0) = c_n e^{-\frac{ig_n A}{\hbar}}$$

which means

$$\psi_n(t) = \psi_n(0)e^{-i\frac{E_n t}{\hbar}}$$

Now, we must check that the state vector remains normalised if our work is to be correct:

$$\langle\psi(t)|\psi(t)\rangle = \left(\sum_m c_m^* e^{\frac{i}{\hbar}(g_m A + E_m t)}\langle u_m|\right)$$

$$\left(\sum_n c_n e^{-\frac{i}{\hbar}(g_n A + E_n t)}|u_n\rangle\right)$$

$$= \sum_{m,n} c_m^* c_n e^{\frac{i}{\hbar}\left[(g_m A + E_m t) - (g_n A + E_n t)\right]}\delta_{mn}$$

$$= \sum_n |c_n|^2 = \langle\psi(0)|\psi(0)\rangle$$

Hence, if we started with a normalised state $\langle\psi(0)|\psi(0)\rangle = 1$, then we are guaranteed that the state will remain normalised.

In fact, Equation (134) guarantees that any evolution equation deduced in this section, like for example, Equation (138), will conserve the normalisation of the state vectors, because according to our deduction, those vectors will all be of the form given by Equation (134); hence,

$$\langle\psi|\psi\rangle = \left(\sum_m c_m^* e^{\frac{i}{\hbar}\theta_m}\langle u_m|\right)\left(\sum_n c_n e^{\frac{i}{\hbar}\theta_n}|u_n\rangle\right)$$

$$= \sum_{m,n} c_m^* c_n e^{\frac{i}{\hbar}(\theta_n - \theta_m)}\delta_{mn}$$

$$= \sum_n |c_n|^2 = \langle\psi_0|\psi_0\rangle$$

where

$$|\psi_0\rangle = \sum_n c_n |u_n\rangle$$

where all the phases in it have been set to zero. We must restress that this whole description is valid not only for quantum experiments but for any probabilistic experiment in general as we have seen so far.

Note: for the sake of completeness, we must say that we could have expanded the state on a basis that represents an experiment with a changing probability distribution as the state changes, and following the same steps as above, we would have gotten the following equation:

$$ib\frac{\partial|\varphi\rangle}{\partial q^j} = \hat{G}_j|\varphi\rangle + ib\sum_m \frac{\partial c_m}{\partial q^j} e^{i\theta_m}|u_m\rangle$$

But it is a very non-elegant equation, which is not very easy to handle. And since we are not forced to do that as has been explained extensively above, we are not going to use this equation any further.

Another note: Notice that we could have chosen another 'time variable' which will give us another Hamiltonian operator with a similar structure of the above. Thus, the choice we made is not unique.

*Discrete evolution*
We can think of the transformation of a general state vector immediately into a state vector after the measurement as discussed above, as a discrete evolution of the state vector. Hence, the measurement, as discussed above, represents a discrete evolution in the state vector.

### 3.1.10 | Complex Hilbert spaces and the quantisation of space-time

In section 3.1.9, we have explained that had we used real Hilbert spaces, we would not have been able to speak of any continuous evolution of states. In particular, we have proved the following:

*If quantum mechanics is described using complex Hilbert spaces, then we can talk about a continuous evolution of states. This is equivalent to saying if we cannot have continuous evolution of states (for example, if space time is quantised) then quantum mechanics is not described using complex Hilbert spaces.*

This means that if experiments like the ones talked about in Ref. [19], which are designed to verify whether we can describe quantum mechanics using real Hilbert space or not, found that we can indeed do so, that indicates that what we perceive in nature as continuous evolution is in reality a discrete one, which means that in this case space-time itself, for example, must be quantised.

Of Course, quantisation of space-time is not only important for quantum gravity [20], but among other things, it is also important for example, in discussions regarding subjects such as *PvsNP* as you can see in Ref. [17, 18]. This makes knowing the answer to the question of whether we can describe quantum mechanics using real Hilbert spaces or not a very interesting and important thing to know.

### 3.1.11 | The expectation value of some observable

Let us take some observable A which can be represented according to section "Observables" by a Hermitian operator in our Hilbert space and let us denote this operator by $\hat{A}$. We know that we can always build an orthonormal basis out of the eigenvectors of $\hat{A}$ (as shown in [15] p. 93), meaning

$$\hat{A}|a_i\rangle = a_i|a_i\rangle, and \langle a_i|a_j\rangle = \delta_{ij}, \text{with} \sum_i |a_i\rangle\langle a_i| = \hat{\mathbb{1}}$$

Now, if the system was in the state $|\psi\rangle$, and if $p_i$ was the probability of getting $a_i$ after the measurement, then

$$<A> = \sum_i a_i p_i = \sum_i a_i \langle a_i|\psi\rangle\langle\psi|a_i\rangle$$
$$= \langle\psi|\left(\sum_i a_i|a_i\rangle\langle a_i|\right)|\psi\rangle$$

Hence,

$$<A> = \langle\psi|\hat{A}|\psi\rangle \tag{145}$$

where we used the spectral composition of $\hat{A}$ ([15] p.119):

$$\hat{A} = \sum_i a_i|a_i\rangle\langle a_i|$$

and since $<A> = \sum_i a_i p_i$ and all $a_i \in \mathbb{R}$ since $\hat{A}$ is Hermitian in a finite-dimensional Hilbert space, and since $p_i \in \mathbb{R}$, this means that $<A> \in \mathbb{R}$.

### 3.1.12 | Doubly generalised Ehrenfest theorem

From Equation (138) and replacing in the notation $|\varphi\rangle$ with $|\psi\rangle$, we can write

$$\frac{\partial|\psi\rangle}{\partial q^j} = -\frac{i}{b}\hat{G}_j|\psi\rangle$$

$$\frac{\partial\langle\psi|}{\partial q^j} = \frac{i}{b}\langle\psi|\hat{G}_j$$

Now, if we take any observable $A$, which is represented by the Hermitian operator $\hat{A}$ we see using Equation (145) that

$$\begin{aligned}\frac{\partial}{\partial q^j}<A> &= \frac{\partial}{\partial q^j}\langle\psi|\hat{A}|\psi\rangle\\ &= \frac{\partial\langle\psi|}{\partial q^j}\left(\hat{A}|\psi\rangle\right) + \langle\psi|\frac{\partial\hat{A}}{\partial q^j}|\psi\rangle + \langle\psi|\hat{A}\left(\frac{\partial|\psi\rangle}{\partial q^j}\right)\\ &= \frac{i}{b}\langle\psi|\left[\hat{G}_j,\hat{A}\right]|\psi\rangle + \langle\psi|\frac{\partial\hat{A}}{\partial q^j}|\psi\rangle\end{aligned}$$

Therefore,

$$\frac{\partial<A>}{\partial q^j} = \frac{i}{b}<\left[\hat{G}_j,\hat{A}\right]> + <\frac{\partial\hat{A}}{\partial q^j}> \qquad (146)$$

which we will call the doubly generalised Ehrenfest theorem. And if we calculate the rate of change with time of the expectation value of some observable we will find, using the doubly generalized Ehrenfest theorem,

$$\begin{aligned}\frac{d}{dt}<A> &= \frac{\partial<A>}{\partial q^j}\frac{\partial q^j}{\partial t}\\ &= \frac{i}{b}<\left[\hat{G}_j,\hat{A}\right]>\frac{\partial q^j}{\partial t} + <\frac{\partial\hat{A}}{\partial q^j}>\frac{\partial q^j}{\partial t}\\ &= \frac{i}{b}\langle\psi|\left[\frac{\partial q^j}{\partial t}\hat{G}_j,\hat{A}\right]|\psi\rangle + \langle\psi|\frac{\partial\hat{A}}{\partial q^j}\frac{\partial q^j}{\partial t}|\psi\rangle\\ &= \frac{i}{b}\langle\psi|\left[\hat{H},\hat{A}\right]|\psi\rangle + \langle\psi|\frac{\partial\hat{A}}{\partial t}|\psi\rangle\end{aligned}$$

which yields,

$$\frac{d}{dt}<A> = \frac{i}{b}<\left[\hat{H},\hat{A}\right]> + <\frac{\partial\hat{A}}{\partial t}> \qquad (147)$$

This is nothing other than generalised Ehrenfest theorem.

### 3.1.13 | The uncertainty principle

If quantum mechanics is just probability theory, then we must find classical counterparts for superposition, the uncertainty principle, and entanglement, without coming to conflict with Bell's theorem. In addition, we have to solve the measurement problem.

And in fact, it has been explained in Ref. [6] how we can find counterparts to superposition, the uncertainty principle, and how this interpretation is in agreement with Bell's theorem, and it solves the measurement problem. But in this paper, we will expand (in addition to what we have laid out already in this work) on the mentioned discussion and make it more quantitative.

We saw in section 3.1.7, that in this formulation, the value that some property takes is defined through the act of measurement according to Equation (107). That means a property is defined to take a certain value only after measurement itself, or in other words, through the act of measurement.

To understand this, let us take the example of a coin toss. What do we mean when we say that the coin is either heads or tails? We can define a unit vector $\hat{C}$ that is perpendicular to the surface of the coin and is pointing from the face that represents tails, to the face that represents heads. The coin is perpendicular to a previously known unit vector (usually it is taken as a normal vector to the surface of the Earth, but not always), and let us denote it by $\hat{n}$, and we say that the coin is heads when $\hat{C}$ and $\hat{n}$ are parallel, while we say that the coin is tails when they are antiparallel, as in the following Figure 1

Thus, the labels heads and tails have no meaning unless we determine relative to what $\hat{n}$ we are assigning them. Here we notice two things: first, we cannot assign the coin to be heads on two different directions; thus, it cannot be *measured* to be heads on more than one direction, and this represents the uncertainty principle in this case. Second, since the result of the throw of the coin is not defined until after it becomes parallel/anti-parallel to some direction after it is thrown, this means that this is in agreement with Bell's theorem, as is explained extensively in Ref. [6], and as will be discussed thoroughly below in the section about entanglement and Bell's theorem. Furthermore, as we have seen above in the example of coin/the person observing the coin, the state of the composite system is an entangled state. Before tossing the coin by someone, we cannot say that the coin was heads nor tails, neither that the observer sees it heads or tails (by sees it we mean that it is assigned some label in their universe).When the person holding the coin throws it and assigns it a specific label relative to some $\hat{n}$, and let us say the label was heads, then immediately in the universe of the observer, we can say that it sees (in the mentioned meaning of the word above) the coin heads. So, nothing mysterious in this case: what is happening in these two entangled systems is that the results of one experiment done on one system are defined using the results of the experiment done on the other.

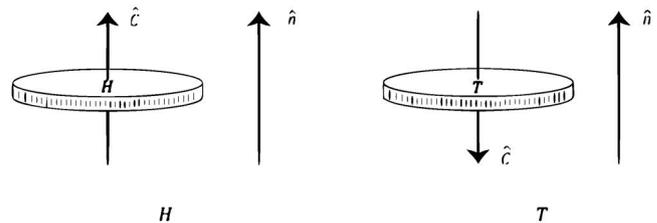

**FIGURE 1** Defining heads and tails

Now, if we take two observables $\hat{A}$ and $\hat{B}$ we can calculate their variances in the state $|\psi\rangle$ according to Equation (145) to be

$$\sigma_A^2 = \left\langle \left(\hat{A} - <A> \mathbb{1}\right)\psi \,\Big|\, \left(\hat{A} - <A> \mathbb{1}\right)\psi \right\rangle$$

and

$$\sigma_B^2 = \left\langle \left(\hat{B} - <B> \mathbb{1}\right)\psi \,\Big|\, \left(\hat{B} - <B> \mathbb{1}\right)\psi \right\rangle$$

Now, since the mathematical deduction of the generalised uncertainty principle in Ref. [15] pp. 108–109 relied solely on the previous mathematical fact and since as we see we retain it in our model, this means that we can repeat the same mathematical steps to get

$$\sigma_A \sigma_B \geq |\frac{1}{2i} < \left[\hat{A}, \hat{B}\right] > | \tag{148}$$

which is the generalised uncertainty principle.

Now, what about energy-time uncertainty principle?

Well, the energy-time uncertainty was deduced in Ref. [15] pp. 112–114 using generalised Ehrenfest theorem, and a definition for $\Delta t$ as the time it takes for the expectation value of a general observable to change by one standard deviation, and the fact that

$$\sigma_A \sigma_B \geq |\frac{1}{2i} < \left[\hat{A}, \hat{B}\right] > |$$

All these facts still hold here; hence, following the same mathematical steps in a manner similar to Ref. [15] pp. 112–114 we find

$$\Delta E \Delta t \geq \frac{b}{2} \tag{149}$$

where $\Delta E = \sigma_H$ and $\Delta t = \frac{\sigma_Q}{|d<Q>/dt|}$, and where we assumed of course that $\frac{\partial \hat{Q}}{\partial t} = \hat{0}$.

Moreover, due to the doubly generalised Ehrenfest theorem we deduced here, we can build a new class of uncertainty relations as follows:

If we take an observable $\hat{A}$ that satisfies

$$\frac{\partial \hat{A}}{\partial q^j} = \hat{0}$$

Then, from doubly generalised Ehrenfest theorem,

$$< \left[\hat{G}_j, \hat{A}\right] > = \frac{b}{i} \frac{\partial}{\partial q^j} <A> \tag{150}$$

which means, using Equation (148) that

$$\sigma_A \sigma_{G_j} \geq \frac{b}{2}|\frac{\partial <A>}{\partial q^j}| \Rightarrow \sigma_{G_j} \frac{\sigma_A}{|\partial <A> / \partial q^j|} \geq \frac{b}{2}$$

If we define $\Delta q^j$ to be

$$\Delta q^j := \frac{\sigma_A}{|\partial <A> / \partial q^j|}$$

and $\Delta G_j$ to be $\Delta G_j = \sigma_{G_j}$ we get

$$\Delta G_j \Delta q^j \geq \frac{b}{2} \quad (no \quad sum) \tag{151}$$

where since

$$\sigma_A = |\frac{\partial <A>}{\partial q^j}| \Delta q^j \quad (no \quad sum)$$

we see that $\Delta q^j$ is the change in $q^j$ that makes $<A>$ change by one standard deviation.

Note:

Due to the existence of gravity in the universe that requires space to be curved, we see that we cannot define $H$ and $T$ for some coin except for a local vector $\hat{n}$ which means for $\hat{n}$ that lies at the same point as the coin (if we assume the coin to be practically a point). This means that if a coin is $H$ according to a coin tosser associated with $\hat{n}$, it may not have a defined result ($H$ or $T$) relative to a distant observer because parallelism is broken between the arrow associated with the latter observer and $\hat{n}$. That means that we may end in some situations that the outcomes themselves can be defined relative to some observers but not for others. This has some similarity to some aspects of relational quantum mechanics, as we can find, for example, in Ref. [21].

### 3.1.14 | Entanglement and Bell's theorem

We have already encountered a special case of entanglement when we talked about the observer-system composite system. But here, we will study more thoroughly entanglement in general. In particular, we will discuss the tests on violation of the Bell type inequalities. It was thoroughly demonstrated in Ref. [16] that the tests on violation of the Bell type inequalities are simply statistical tests of local incompatibility of observables using only the known mathematical structure of quantum mechanics. And since we have deduced the full mathematical structure of quantum mechanics and proved that it is also valid for classical systems, this means that the same arguments made in Ref. [16] still hold here. In particular what differs here from the work in Ref. [16] is that in Ref. [16] the local incompatibility of observables (manifested in the non-commutativity of their operators) was treated as a purely quantum phenomena, where in this work we see that it also applies to classical systems, and we will see this incompatibility in detail in the following example.

## 3.1.15 | An example that wraps everything up

We will describe one particular classical probabilistic experiment and show that it has every feature we have in quantum mechanics.

Let us take an ensemble of unit length arrows with each arrow having a mark on it, which takes a random position (meaning the positions of the marks on the arrows are distributed randomly and uniformly) on it, and we will denote the $i$th arrow in the ensemble by the symbol $\hat{c}_i$.

Now, we have a certain device that has a unit length arrow attached to it, and let us call it $\hat{n}$.

The measurement happens as follows: The measurement device takes an arrow from the ensemble randomly and identify the half point of $\hat{n}$ with the half point of $\hat{c}_i$, and then it projects $\hat{n}$ onto $\hat{c}_i$. If the mark on $\hat{c}_i$ lies within this projection, the device aligns $\hat{c}_i$ with $\hat{n}$, and then registers the value $H$. Otherwise, it anti-aligns $\hat{c}_i$ with $\hat{n}$ and then registers $T$ as we can see in Figure 2.

If all $\hat{c}_i$s make the angle $\theta$ with $\hat{n}$, then from the Figure 2 we see that the probability of getting $H$ will be $1/2 + (1/2\cos\theta) = \cos^2\frac{\theta}{2}$.

This means that for the previous ensemble, the probability of getting $H$ along an arbitrary unit vector in space will be $\cos^2\frac{\theta}{2}$, where $\theta$ is the angle that $\hat{c}_i$s make with $\hat{n}$.

We notice that if we did the measurement and got, for example, $H$ on some direction, next we redo the same measurement, and we will get the same result. We see that the probability distribution is identical to the probability distribution of measuring the spin of an electron along some axis. This means that we can use a similar representation to the Bloch sphere representation to write the state vector for measuring $H$ and $T$ for the arrows.

$$|\psi\rangle = \cos\frac{\theta}{2}|H\rangle + e^{i\phi}\sin\frac{\theta}{2}|T\rangle \qquad (152)$$

This means that the eigenvectors that represent measuring the arrows to be $H$ and $T$ on the $z$-axis are $|H\rangle, |T\rangle$, respectively. While the ones representing measuring the arrows to be $H$ and $T$ along the $x$-axis are $\frac{1}{\sqrt{2}}(|H\rangle + |T\rangle), \frac{1}{\sqrt{2}}(|H\rangle - |T\rangle)$, respectively. And the eigenvectors that represent measuring the arrows to be $H$ and $T$ along the $y$-axis are $\frac{1}{\sqrt{2}}(|H\rangle + i|T\rangle)$, and $\frac{1}{\sqrt{2}}(|H\rangle - i|T\rangle)$, respectively.

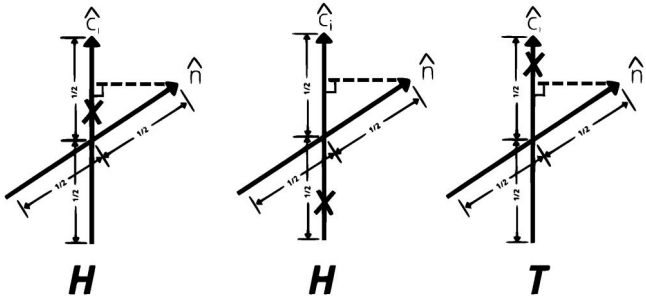

**FIGURE 2**  Measuring $H$ and $T$ along some direction

We can, for example, build the following three Hermitian operators, in a similar manner to the method layed out in section "Observables" for some positive real number $\alpha$:

$$\hat{G}_3|H\rangle = \alpha|H\rangle$$

$$\hat{G}_3|T\rangle = -\alpha|T\rangle$$

And

$$\hat{G}_1\left(\frac{|H\rangle + |T\rangle}{\sqrt{2}}\right) = \alpha\left(\frac{|H\rangle + |T\rangle}{\sqrt{2}}\right)$$

$$\hat{G}_1\left(\frac{|H\rangle - |T\rangle}{\sqrt{2}}\right) = -\alpha\left(\frac{|H\rangle - |T\rangle}{\sqrt{2}}\right)$$

Finally,

$$\hat{G}_2\left(\frac{|H\rangle + i|T\rangle}{\sqrt{2}}\right) = \alpha\left(\frac{|H\rangle + i|T\rangle}{\sqrt{2}}\right)$$

$$\hat{G}_2\left(\frac{|H\rangle - i|T\rangle}{\sqrt{2}}\right) = -\alpha\left(\frac{|H\rangle - i|T\rangle}{\sqrt{2}}\right)$$

We will represent them collectively as,

$$\hat{\vec{G}} = \mathbf{e_1}\hat{G}_1 + \mathbf{e_2}\hat{G}_2 + \mathbf{e_3}\hat{G}_3$$

where $\mathbf{e_1}$, $\mathbf{e_2}$, and $\mathbf{e_3}$ are the unit vectors along the $x$, $y$, and $z$ axes, respectively.

We notice that $\hat{G}_3$ represents an observable that can be measured when $\hat{n} = \mathbf{e_3}$ and this observable will take one of two values: $\alpha$ or $-\alpha$. The same can be said about $\hat{G}_1$ and $\hat{G}_2$ but along the other coordinate axes.

From how we defined the measurement in this experiment, we immediately see that we cannot measure any two of the previous observables simultaneously.

Now the matrices representing the previous observables in the basis $\{|H\rangle, |T\rangle\}$ will be

$$G_1 = \begin{pmatrix} 0 & \alpha \\ \alpha & 0 \end{pmatrix} = \alpha\sigma_1$$

$$G_2 = \begin{pmatrix} 0 & -i\alpha \\ i\alpha & 0 \end{pmatrix} = \alpha\sigma_2$$

$$G_3 = \begin{pmatrix} \alpha & 0 \\ 0 & -\alpha \end{pmatrix} = \alpha\sigma_3$$

where $\sigma_1$, $\sigma_2$ and $\sigma_3$ are the three Pauli matrices.

It is a very nice result to get the three Pauli matrices (which they naturally spring from Dirac equation, a relativistic quantum mechanical equation) from a simple classical probabilistic experiment using our mathematics as this experiment.

From the above matrices, we can easily deduce the commutation and anti-commutation relations between those observables:

$$[G_a, G_b] = 2i\varepsilon_{abc}\alpha^2\sigma_c = 2i\varepsilon_{abc}\alpha G_c \quad (no \quad sum) \quad (153)$$

$$\{G_a, G_b\} = 2\delta_{ab}\alpha^2 I \quad (154)$$

where $\varepsilon_{abc}$ is the Levi-Civita symbol, $\delta_{ab}$ is the Kronecker delta, and $I$ is the identity matrix.

And since $\hat{G}_1$, $\hat{G}_2$, and $\hat{G}_3$ do not commute, this is with total agreement with what we said about the uncertainty principle.

Moreover, when $a \neq b$ we find from Equation (148) and Equation (153) that

$$\sigma_{G_a}\sigma_{G_b} \geq \alpha|<G_c>| \quad (155)$$

Now, we will take a particular kind of ensemble. We will take the ensemble of arrows that have the same direction and lie in the xy-plane and that all of them are rotating around the z-axis with a constant angular velocity $\omega$.

For this ensemble, the probability of measuring $H$ or $T$ along the z-axis (along $\mathbf{e}_3$) is $\cos^2\left(\frac{\pi/2}{2}\right) = \frac{1}{2}$, and these probabilities will not change, which means that we can expand the state vector on the eigenvectors of this experiment when we want to use Equation (143) to study the evolution of the state of the system.

Thus, the state vector for this ensemble can be written according to Equation (152) as

$$|\psi\rangle = \frac{1}{\sqrt{2}}|H\rangle + \frac{1}{\sqrt{2}}e^{i\phi}|T\rangle \quad (156)$$

On the other hand, according to Equation (134), we should be able to write the previous state also in the following form:

$$|\psi\rangle = \frac{1}{\sqrt{2}}e^{i\phi_1}|H\rangle + \frac{1}{\sqrt{2}}e^{i\phi_2}|T\rangle$$

which can be written according to Equation (137) as

$$\frac{1}{\sqrt{2}}e^{-i\alpha\frac{f(\phi)}{b}}|H\rangle + \frac{1}{\sqrt{2}}e^{i\alpha\frac{f(\phi)}{b}}|T\rangle$$
$$= e^{-i\alpha\frac{f(\phi)}{b}}\left(\frac{1}{\sqrt{2}}|H\rangle + \frac{1}{\sqrt{2}}e^{i\frac{2\alpha f(\phi)}{b}}|T\rangle\right)$$

And since $e^{-i\alpha\frac{f(\phi)}{b}}$ is merely a pure phase, that means we can write the state as

$$|\psi\rangle = \frac{1}{\sqrt{2}}|H\rangle + \frac{1}{\sqrt{2}}e^{i\frac{2\alpha f(\phi)}{b}}|T\rangle \quad (157)$$

Comparing Equation (156) with Equation (157), we can write

$$\frac{2\alpha f(\phi)}{b} = \phi$$

which means that

$$f(\phi) = \frac{b\phi}{2\alpha} \quad (158)$$

But since the arrows rotate with the constant angular velocity $\omega$, we can write

$$\phi = \omega t$$

Now, we can build the Hamiltonian in the following manner. We will define the following operator:

$$\hat{G}_\phi = \frac{\partial f}{\partial\phi}\hat{G}_3 = \frac{b}{2\alpha}\hat{G}_3$$

Hence, according to Equation (139) and Equation (142), the Hamiltonian will be

$$\hat{H} = \frac{\partial\phi}{\partial t}\hat{G}_\phi$$

or

$$\hat{H} = \frac{b\omega}{2\alpha}\hat{G}_3 \quad (159)$$

And its matrix will be

$$H = \frac{b\omega}{2\alpha}\begin{pmatrix} \alpha & 0 \\ 0 & -\alpha \end{pmatrix}$$
$$= \frac{b\omega}{2}\begin{pmatrix} 1 & 0 \\ 0 & -1 \end{pmatrix}$$

That means

$$H = \frac{b\omega}{2}\sigma_3 \quad (160)$$

We can write Equation (159), using the definition we adopted for $\overrightarrow{\hat{G}}$ as

$$\hat{H} = \frac{b\omega}{2\alpha}\overrightarrow{\hat{G}}.\mathbf{e_3} \quad (161)$$

which is extremely analogous to the Hamiltonian of an electron in a magnetic field along $\mathbf{e_3}$, especially when we remember that $b$ itself is nothing but Planck's constant.

## 3.2 | Infinite dimensional outcome spaces

Since we have established the Hilbert space structure needed in quantum mechanics for finite dimensional cases, we can make it our starting point and extend it in the same manner done in Ref. [6] to the infinite dimensional case. This is why we will not delve more into this matter in the current paper.

## 4 | MANAGERIAL INSIGHTS

As will be discussed thoroughly in the next two sections, this paper paves the way to new paths of research that may open the door to build more reliable types of qbits, which would help making quantum computing more practical and cheap.

## 5 | CONCLUSION

The result of this work is that quantum mechanics is just probability theory cast in another language.

This leads to the following conclusions:

1. Usually, books on probability theory define the sample space as the set of all possible outcomes. But outcomes themselves are not well defined, neither the concept of probabilistic experiment. What was done in this work is that through a rigorous treatment of the previous concepts using category theory, we can clearly see that the laws of probability theory are nothing other than the laws of quantum mechanics itself, and the superficial distinction between the two should be removed.

2. We do not have to assume the existence of infinite number of universes in order to answer the question of why the laws of physics are the way they are as Max Tegmark has done [10]. Because the answer to this question will simply be 'because the universe is non-deterministic, thus it will be described using probability theory, hence quantum mechanics'. On a personal level, I find this answer extremely philosophically pleasing. Nonetheless, I still prefer to look if the laws of quantum mechanics are the statistical description of a deeper deterministic layer of reality in the same sense that probability theory is a valid description for a Newtonian universe if we do not have enough information about the system that is under study. Finally, even if the theory that describes the universe is deterministic, it can be described using this structure as we have shown above in Section 3.
   In fact, we can go even further: Maybe there is no physical laws that govern the particles in the universe and their behaviour is completely random. But because probability theory can be applied to study randomness, then those particles can be described using probability theory; hence, our universe will be governed by the laws of quantum mechanics.

3. Since we can use probability theory to describe 'classical systems', thus, classical systems can be described using this structure of quantum mechanics but in this new understanding; hence, we can apply quantum algorithms to them, which gives us more freedom in choosing the components out of which we can build quantum computers, which may give us new paths that make it more practical.

4. Since we can now simulate a quantum computer using a classical system (in principle at least), what this work reveals is that building algorithms using this probabilistic structure is more efficient than the conventional algorithms. This is also why the question of why our universe is quantum mechanical rather than classical is misplaced because the only difference that exists between the classical case and the quantum mechanical one lies in the way we define observables and in which observables are compatible and which are not in each case, as we have explained in section "Observables".

5. What this paper shows regarding the discussion of determinacy versus non-determinacy of measured physical properties such as position, for example, is that even if there is some deterministic level from which our universe emerges, that does not necessarily mean there are some hidden variables which if we know, will make the position deterministic. Because the position-as a result of a probabilistic experiment-is in itself a part of a statistical description and what might be deterministic in such a case are different sets of degrees of freedom that are different from the physical properties we usually measure.

6. Since we saw that classical systems described using ordinary probability theory as formulated in this paper can reproduce the whole mathematical structure of quantum mechanics, which means that we should be able to imitate the behaviour of quantum systems using classical ones, this gives more weight to the loopholes in Bell tests, if we wanted to investigate the idea of finding a deeper deterministic level of reality, which means that we should investigate them more thoroughly, especially since no experiment to date is totally loophole-free [11]. Not to mention that we already have working local hidden variable models that produce the same singlet correlations that we get experimentally as you can find for example, in Ref. [12–14].

## 6 | WHICH IS MORE FUNDAMENTAL: PHYSICS OR MATHEMATICS?

It is a very surprising result that one can through math alone, deduce the laws of physics (in this case, the laws of quantum mechanics). This fuels again the discussion about which is more fundamental: math or physics? Before we even start trying to answer, let us make some definitions for some of the most basic concepts.

Let us define a true claim as the following:

*We say that a claim about a certain aspect of existence is true if this aspect of existence works in the same way described by this claim. Otherwise, we say that the claim is false.*

Thus, it is obvious that to verify whether a claim is true or not, we need to make an experiment to test it (to compare it to reality).

If we want to construct a model that describes the world around us, not just any model, but the most reliable model we can come up with, this model must be built only upon verified claims, which means that all the claims in this model had to have been experimentally verified, and this in a nutshell is the spirit of the scientific enterprise.

Of course, from the previous discussion, we see that claims which cannot, even in principle, be experimentally verified, will be of no use in building such a model; thus, they will be of no interest to us here, regardless whether they are true or not.

But what about logic and mathematics?

Let us give a somewhat new definition for logic (it is new as far as I know): *Logic is the set of general rules that govern the behaviours of things in the world (or some of their behaviours), regardless of the nature of those things themselves.*

For example, when we say that an object must be ($A$) or not ($A$), we mean that this is true whatever that object was: any object is either a pen or not a pen, the same is true if we have used instead of a pen, a table or any other object for that matter. And we have made the previous assertions based upon our observation and experiment (similar arguments work for other types of logic, say fuzzy logic, for example: their rules are taken from the descriptions of some aspects of reality). In fact, I will go too far in this matter as to state my strong conviction that logic itself is something which is deeply rooted in experiment. And since the mathematics we use is deeply rooted in logic, this explains its seemingly unbelievable effectiveness in describing the world. In fact, since this mathematics is based on logic, which is the general rules that govern things regardless of their nature, it is then of no great surprise that mathematics is able to describe the behaviour of elementary particles, even though we cannot imagine those particles or construct any visual picture for them. Our imagination is restricted by two factors: our evolutionary history, which shaped our brains and their ability to build models, and our experience during our lifetime, which supplies us with pictures we can use to construct such models. But neither in our evolutionary history, nor in our daily experience, we have experienced directly the realm of elementary particles. Nonetheless, the success of elementary particle physics, which uses mathematics to describe these particles, tells us that these particles are governed, at least to some extent, by the same general rules that govern our daily life objects.

This however raises the troubling question: are there levels of existence so alien to us that they are not governed by any of the general rules we have been talking about? If that is the case, then we probably will not be able to know about them, but that is a matter of philosophy not of science; this is why this question will concern us here no more (the other possibility I am troubled with is to exist more than one mathematical model that are all able to describe our universe equally well, but they provide radically different ways to interpret reality. But this question is easier than the previous one, because it is still a question of mathematics and physics).

In conclusion, we summarise the characteristics, which we seek in our models that describe reality in order to make them the most reliable models possible:

They must only consist of experimentally verified claims and have a strong logical structure, which means following logic rules in building them.

Now, we have said that we can apparently deduce the laws of physics (in particular, quantum mechanics) using mathematics alone. But if the above discussion was right, then mathematics itself is something which is deeply rooted in experiment. This means that the lines between physics and mathematics are really blurry and maybe the distinction between them is superficial after all.

# 7 | FUTURE RESEARCH SCOPE

As we have seen in the Conclusion section, this work shows that we can, at least in principle, imitate quantum systems using classical ones. This in itself opens new opportunities to start building new types of qbits that are more reliable and easier to control. This work also shows us that we should investigate more thoroughly the loopholes in Bell tests (taking into account that in this model, the result of a measurement is defined only after the measurement, which gives us more ways to investigate those loopholes), to investigate whether there is a deeper deterministic layer of reality or not, especially since some of these loopholes really show us a way to get the correlations expected from quantum mechanics as we have explained in the Conclusion section. Furthermore, this work shows that we should do the needed experiments to know whether we can describe quantum mechanics using real Hilbert spaces or not, because it will have great impact upon our knowledge regarding whether space-time is quantised or not, as we have explained extensively in Section 3.1.10.

## ACKNOWLEDGEMENTS
I wish to express my gratitude for my colleagues Theophanes Raptis, Osama Karkout, and Anwar Alameddin, for their invaluable comments, feedback, and insights throughout the many discussions we had. I would also like to thank my friend Hussain Al Sneeh for his great help in typing and editing this document.

## CONFLICT OF INTEREST
The authors declare no conflicts of interest.

## DATA AVAILABILITY STATEMENT
Data sharing not applicable to this article as no datasets were generated or analysed during the current study.

## ORCID
*Raed Shaiia*  https://orcid.org/0000-0003-2440-8740